\documentclass[11pt,a4paper]{article}

\usepackage{jheppub} 

\usepackage[utf8]{inputenc}
\usepackage{graphicx,dcolumn,subcaption,mathtools}

\usepackage{braket}
\usepackage{siunitx}
\usepackage{appendix}
\usepackage[mathscr]{eucal}
\usepackage{bbm}
\usepackage{cancel}

\usepackage{footnotebackref}

\usepackage{rotating}

\usepackage[]{float}
\usepackage[font=small, labelfont=bf]{caption}

\usepackage{slantsc}
\usepackage{xspace}

\usepackage{setspace}
\allowdisplaybreaks

\newcommand\asbare{\alpha_{\mathrm{S}}^u} 
\newcommand\as{\alpha_{\mathrm{S}}(\mu_R)}

\newcommand\asopi{\f{\as}{\pi}}

\newcommand\f[2]{\frac{#1}{#2}} 

\newcommand{\DeltaET}{\Delta E_t}
\newcommand{\ktness}{k_T^{\text{ness}}}

\newcommand{\kpmunu}{\hat{k}_T^\mu \hat{k}_T^\nu}
\newcommand{\NLO}{\text{NLO}}

\newcommand{\M}{\mathcal{M}}

\newcommand{\T}{\mathcal{T}}
\newcommand{\A}{\mathcal{A}}

\newcommand{\I}{\mathcal{I}}
\renewcommand{\P}{\mathcal{P}}
\newcommand{\E}{\mathcal{E}}

\newcommand{\Amunu}{ \A^{\mu\nu}}

\newcommand{\ISC}{\mathcal{I}^c}
\newcommand{\FSC}{\mathcal{F}^c}
\newcommand{\SOFT}{\mathcal{S}^{wa}}
\newcommand{\VIRT}{\mathcal{V}}
\newcommand{\CTMSB}{\mathcal{C}_{\overline{\text{MS}}} }

\newcommand{\BSigma}{\mathbf{\Sigma}}
\newcommand{\BT}{\mathbf{T}}

\newcommand{\pjperp}{p_{J,t}}

\newcommand{\lpjetbar}{E_J}

\newcommand{\tp}{\tilde{p}}

\newcommand{\rcut}{r_{\mathrm{cut}}}
\newcommand{\DETISC}{\Delta E_t^{\ISC}}
\newcommand{\DETFSC}{\Delta E_t^{\FSC}}

\newcommand{\sub}{\rm sub}

\newcommand{\Matrix}{{\sc Matrix}\xspace}
\newcommand\OpenLoops{{\sc OpenLoops}\xspace}
\newcommand\Recola{{\sc Recola}\xspace}

\title{
Exploring slicing variables for jet processes
}
\preprint{
  \begin{flushright}
    ZU-TH 36/23\\
  \end{flushright}
}

\author[a]{Luca Buonocore,}
\author[a]{Massimiliano Grazzini,}
\author[a]{J\"urg Haag,}
\author[a]{Luca Rottoli}
\author[a]{and Chiara Savoini }

\affiliation[a]{Physik Institut, Universit\"at Z\"urich, CH-8057 Z\"urich, Switzerland}

\emailAdd{lbuono@physik.uzh.ch}
\emailAdd{juerg.haag@physik.uzh.ch}
\emailAdd{grazzini@physik.uzh.ch}
\emailAdd{luca.rottoli@physik.uzh.ch}
\emailAdd{csavoi@physik.uzh.ch}

\date{Received: date / Accepted: \today}

\abstract{
We consider the class of inclusive hadron collider processes in which one or more energetic jets are produced, possibly accompanied by colourless particles. We provide
a general formulation of a slicing scheme for this class of processes, by identifying the various contributions that need to be computed up to next-to-leading order (NLO) in QCD perturbation theory.
We focus on two novel  observables, the one-jet resolution variable $\Delta E_t$ and the $n$-jet resolution variable $\ktness$, and explicitly compute all the
ingredients needed to carry out NLO computations using these variables. We contrast the behaviour of these variables when the slicing parameter becomes small. In the case of $\ktness$ we also present results for the hadroproduction of multiple jets.
}

\arxivnumber{2307.11570}

\begin{document}

\maketitle
\flushbottom

\section{Introduction}
\label{sec:intro}

Processes featuring multiple jets play a crucial role at hadron colliders.
Many important new-physics signatures are characterised by multijet final states plus additional colourless particles.
The evaluation of the corresponding cross sections and kinematical distributions in perturbative QCD requires the availability of the corresponding scattering amplitudes,
and efficient methods to handle and cancel the associated infrared (IR) singularities.
The required scattering amplitudes at tree level and one-loop can nowadays be obtained with automated tools, while two-loop amplitudes are available only for relatively simple processes (see e.g. Refs.~\cite{Heinrich:2020ybq,Huss:2022ful} and references therein).
Several methods to handle and cancel IR singularities have been developed and used to obtain perturbative QCD predictions at the next-to-next-to-leading order (NNLO) for many benchmark processes (see e.g. Ref.~\cite{TorresBobadilla:2020ekr}).
Despite this, the availability and reproducibility of differential NNLO predictions by using public numerical programs is still limited to relatively simple processes \cite{Gavin:2010az,Grazzini:2017mhc,Camarda:2019zyx,Catani:2019hip,Campbell:2019dru,Campbell:2022gdq}.

{\it Non-local} subtraction or {\it slicing} {\footnote{The slicing method was introduced in the context of NLO calculations, first for $e^+e^-$ annihilation~\cite{Fabricius:1981sx,Kramer:1986mc,Giele:1991vf} and, later, for hadron collisions~\cite{Giele:1993dj}. } methods \cite{Catani:2007vq,Stewart:2010tn}
have provided very efficient ways to obtain NNLO predictions for a number of benchmark hadron collider processes involving colourless final states \cite{Grazzini:2008tf,Catani:2009sm,Ferrera:2011bk,Ferrera:2014lca,Ferrera:2017zex,deFlorian:2016uhr,Catani:2011qz,Grazzini:2013bna,Grazzini:2015nwa,Cascioli:2014yka,Grazzini:2015hta,Gehrmann:2014fva,Grazzini:2016ctr,Grazzini:2016swo,Grazzini:2017ckn,Gaunt:2015pea,Boughezal:2016wmq,Campbell:2016yrh,Heinrich:2017bvg,Campbell:2017aul,Catani:2018krb,Abreu:2022zgo}, possibly accompanied by one jet \cite{Boughezal:2015dva,Boughezal:2015aha,Boughezal:2015ded,Boughezal:2016dtm}, and/or a heavy-quark pair \cite{Catani:2019iny,Catani:2020kkl,Catani:2019hip,Catani:2022mfv,Buonocore:2022pqq,Buonocore:2023ljm}. For the simplest processes even next-to-next-to-next-to leading order (N$^3$LO) results have been obtained \cite{Cieri:2018oms,Camarda:2021ict,Billis:2021ecs,Chen:2022cgv,Neumann:2022lft,Chen:2022lwc} with such methods.

Slicing methods are based on identifying a resolution variable to distinguish configurations in which one or more additional QCD partons are resolved.
The resolution variable is then used to introduce a cut in the phase space: the contribution below the cut can be approximated by exploiting the knowledge of the IR behaviour of the corresponding QCD matrix elements, while the contribution above the cut necessarily involves at least one additional parton and can be evaluated by performing a lower order computation.
In the case of the production of a colourless final state and/or heavy quarks a well established resolution variable is the transverse momentum $q_T$ of the triggered final state.
In the case of multijet production a well-known resolution variable is $N$-jettiness \cite{Stewart:2010tn}, $\tau_N$,
which is defined on events containing at least $N$ hard jets.
Requiring $\tau_N\ll 1$ effectively provides an inclusive way to veto additional jets.
Besides their applications as slicing variables, both $q_T$ and $N$-jettiness have also been used as resolution variables in
  matching NNLO calculations to Monte Carlo parton showers \cite{Alioli:2013hqa,Hoche:2014dla,Monni:2019whf,Mazzitelli:2020jio,Alioli:2021qbf}.
  Further examples of resolution variables in hadron collisions are provided by {\it shape variables} \cite{Banfi:2004nk,Banfi:2010xy},
    which are designed to measure the deviation from the leading order (LO) energy flow.
      
The advantage of slicing methods is in the fact that the cross section above the cut can be carried out in a simple way,
and at NNLO it is obtained through well established local NLO subtraction schemes \cite{Catani:1996jh,Catani:1996vz,Catani:2002hc,Frixione:1995ms,Frixione:1997np}. The price to pay is that the approximation of the cross section below the cut introduces a dependence on the slicing parameter. This dependence leads to missing power suppressed contributions and the exact result can be recovered only through a suitable extrapolation procedure.

There are several features that characterise a resolution variable. The process (in)de\-pendence, the factorisation properties
in the IR limit (and the related possibility to carry out all-order resummation),
the absence of non-global logarithmic contributions \cite{Dasgupta:2001sh},
the qualitative behaviour and the quantitative impact of the power suppressed contributions
are all important aspects to establish the extent to which a resolution variable can be useful.
These aspects are in turn relevant also when the variable is used
in the matching of fixed-order calculations to Monte Carlo parton showers. Therefore, the exploration of new resolution variables is interesting by itself, both in the context of fixed-order calculations and of Monte Carlo generators.

In this paper we consider a rather general class of processes: the hadronic production of an arbitrary number of jets, possibly accompanied by a colourless system $F$. We start by providing a general formulation of a slicing scheme for this class of processes, by identifying the various contributions that need to be computed at NLO.
These contributions correspond to what in Soft Collinear Effective Field Theory (SCET)~\cite{Bauer:2000yr,Bauer:2001yt,Bauer:2002nz,Beneke:2002ph,Beneke:2002ni} are called parton level {\it beam}, {\it jet} and {\it soft} functions.
We then focus on two novel observables, the one-jet resolution variable  $\Delta E_t$ and the $n$-jet resolution variable $\ktness$ \cite{Buonocore:2022mle}, and explicitly evaluate all the contributions necessary to implement an NLO computation by using these variables. We also present numerical results by contrasting the different power suppressed contributions affecting the two variables.

The paper is organised as follows. In Sect.~\ref{sec:slicing} we discuss the general structure of the NLO computation for an arbitrary resolution variable.
In Sect.~\ref{sec:applications} we consider two specific examples: in \ref{sec:DeltaET} we focus on the one-jet resolution $\Delta E_t$ variable for the $F+{\rm jet}$ process, while in Sect.~\ref{sec:ktness} we move to the $\ktness$ variable. 
Our numerical results are presented in Sect.~\ref{sec:results}. More details on the computations and analytical results are presented in the Appendices.

\section{Slicing at NLO for jet processes}
\label{sec:slicing}

\subsection{Generalities}
\label{sec:below_cut_general}
In this work we consider processes in which $n$ hard jets are produced, possibly in association with a colourless system $F$.
At the Born level, the kinematics of this process is fully determined by the momentum of the colourless system $p_F$ and the momenta $p_1,p_2,p_3,\dots,p_{n+2}$ of $n+2$ hard QCD massless partons
\begin{equation}
	a_1(p_1) + a_2(p_2) \to a_3(p_3) +\dots + a_{n+2}(p_{n+2}) +  F(p_F) \,,
\end{equation}
where $a_1$, $a_2$,..., $a_{n+2}$ denote the parton flavours. 
Momentum conservation implies
\begin{equation}
	p_1+p_2=p_3+...p_{n+2}+p_F\equiv q.
\end{equation}
At NLO, we also have to consider real configurations with an additional unresolved parton with momentum $k$.
In our notation, momenta labeled with Greek indices $\alpha=1,2...n+2$ refer to all coloured massless Born level partons, while momenta labeled with Latin indices $i=3...n+2$ are associated with final-state partons.
Concerning the flavour of a given QCD parton, we introduce a calligraphic capital letter $\A$ to define a multiindex describing a certain Born channel. For instance $\A=\{a_\alpha\}=\{a_1,a_2,\{a_i\}\}$ would refer to the channel $a_1+a_2\to \{a_i\}+F$.

At NLO, a slicing method based on a resolution variable $r$ (that we assume to be properly normalised to make it dimensionless) is in general built by splitting the hadronic cross section into a contribution above and a contribution below a small cut $\rcut$
\begin{align}
	\sigma^\NLO& = \int_{n+1} d\sigma^R +  \int_n (d\sigma^V+ d\sigma^B)\notag\\
	&=  \int_{n+1} d\sigma^R\Theta(r-\rcut) +  \left(\int_{n+1} d\sigma^R\Theta(\rcut-r) +   \int_n (d\sigma^V+ d\sigma^B)\right)\, ,
\end{align}
where $ d\sigma^B $and $d\sigma^R$ are the Born and real emission contributions respectively, while $d\sigma^V$ contains the genuine loop-tree interference diagrams and the mass factorisation counterterms
\begin{align}
	\label{eq:Virt_below}
	\int_n d\sigma^V =\VIRT + \CTMSB\, .
\end{align}
The contribution above the cut is IR-finite in $d=4$ dimensions and it can be integrated numerically with a Monte Carlo method. The calculation of the contribution below the cut can be carried out in an analytic fashion by approximating the phase space, the resolution variable and the real matrix element in the relevant IR limits. The integration needs to be performed in $d=4-2\epsilon$ dimensions and the IR poles from the real integration cancel the explicit poles from $d\sigma^V$.
Throughout this paper we work in the conventional dimensional regularisation (CDR) scheme, with two polarisations for massless \mbox{(anti-)quarks} and $d-2$ polarisations for gluons. The strong coupling $\as$ is renormalised in the $\overline{\text{MS}}$-scheme and related to the bare coupling $\asbare$ via
\begin{align}
  	\asbare \mu_0^{2 \epsilon} S_\epsilon=\as \mu_R^{2 \epsilon}\left[1-\frac{\as}{\pi}\frac{\beta_0}{\epsilon}+\mathcal{O}(\alpha^2_{\rm S})\right] \,,
\end{align}
where
$S_\epsilon=(4\pi)^\epsilon e^{-\gamma_E \epsilon}$, $\beta_0=11C_A/12-T_R n_f/3$ and $\mu_R$ is the renormalisation scale.
The $SU(N_c)$ QCD colour factors are $C_F=(N_c^2-1)/(2N_c)$, $C_A=N_c$, $T_R=1/2$ and $n_f$ is the number of massless flavours.
In general we can write the contribution below the cut as
\begin{align}
\label{eq:below_cut_coefs}
  & \!\int_{n+1} \!\!\! d\sigma^R\Theta(\rcut-r) +   \!\!\int_n (d\sigma^V+ d\sigma^B)  \notag \\
  & =
\!\!\!\! \sum_{\A,\,\{b_1,b_2\}}\! \int_0^1 \! dx_1  \!\int_0^1 \! dx_2\!\int_{x_1}^1\frac{dz_1}{z_1}\int_{x_2}^1\frac{dz_2}{z_2}
  f_{b_1}\!\left(\frac{x_1}{z_1},\mu_F\right)\!f_{b_2}\!\left(\frac{x_2}{z_2},\mu_F\right)  \notag \\
  &\!\!\times\!\!\!\int\!\! \frac{d\Phi_B}{2Q^2}\bra{ \M^{(0)}_{\A} }\!\! \left[ \BSigma^{00}_{\A b_1 b_2}(z_1,z_2)+\asopi \Biggl( \sum_{k=0}^2 \BSigma^{1k}_{\A b_1 b_2}(z_1,z_2)\log^k(\rcut)+{\cal O}(\rcut^p)\Biggr)\right]\!\! \ket{ \M^{(0)}_{\A}},
\end{align}
where $d\Phi_B$ is the four-dimensional Born phase space, $Q = \sqrt{q^2}$ is the invariant mass of the Born event
and $\ket{\M^{(0)}_{\A}}$ is the Born matrix element (which can be evaluated here in $d=4$ dimensions),
with $\ket{\cdot}$ denoting a vector in colour space (see e.g. Ref.~\cite{Catani:1996vz}).
In the above formula, $f_{a}(x,\mu_{F})$ is the parton distribution function (PDF) of parton $a$ carrying a fraction $x$ of the proton momentum at the factorisation scale $\mu_{F}$.
Bold symbols denote operators acting on colour space and we defined
\begin{equation}
  \BSigma^{00}_{\A b_1 b_2}(z_1,z_2) =
   \mathbf{1}
  \delta_{a_1 b_1}\delta_{a_2 b_2}\delta(1-z_1) \delta(1-z_2) \,,
\end{equation}
where $\mathbf{1}$ is the identity operator in colour space.
We anticipate that in Eq. (\ref{eq:below_cut_coefs}) the missing power corrections in $\rcut$ can be logarithmically enhanced for some slicing variables.

The complicated part of the calculation is the integration of the real emission contribution over the $1$-particle radiation phase space subjected to the constraint $r<\rcut$, retaining the full dependence on the Born kinematics. Indeed, in this region, the integral is dominated by configurations in which a parton is soft and/or is radiated collinearly to one of the $n+2$ external legs. In order to extract the leading power behaviour in $\rcut$, our strategy is based on approximating both the real matrix element squared, using the factorisation properties of QCD tree-level amplitudes, and the observable $r$ in the relevant IR limits. The treatment of the phase space requires some care. Indeed, a naive approximation of the phase space in the different limits may lead to integrals that are divergent in $d$ dimensions. This is the well-known problem of rapidity divergences~\cite{Collins:1992tv,BenekeFA,Collins:2011zzd} occurring in approaches based on the method of regions~\cite{Beneke:1997zp,Smirnov:2002pj}, such as 
SCET.

In the following we will detail the construction of suitable approximations to obtain the analytic expression, at leading power, for the real emission contribution below the cut. To achieve this result, it is natural to organise the calculation by separating the IR singular regions as
\begin{align}
	\label{eq:Real_below}
	\int_{n+1} d\sigma^R\Theta(\rcut-r) = \ISC+\FSC+\SOFT \, ,
\end{align}
i.e. as a sum of initial-state collinear contributions, $\ISC$, final-state collinear contributions, $\FSC$, and a soft one, $\SOFT$. These three quantities must be properly defined to avoid the double counting in the soft-collinear regions. In our approach, we retain the relevant soft-collinear configurations in $\ISC$ and $\FSC$ and include the left-over soft wide-angle emissions in $\SOFT$, thus avoiding double counting.
The resulting ingredients lead to the definition of perturbative beam, jet and soft functions, following the nomenclature used in the SCET literature~\footnote{Notice that our definitions may differ from those customarily used in SCET. See Appendix~\ref{app:Scet-like-jet-function} for further details on the comparison between the SCET jet function and the one defined in this work.}.

\subsection{Initial-state collinear limit}
\label{sec: ISC}
In this Section we outline the computation of the initial-state collinear contribution in the region below the cut, $r<r_{\rm cut}$,
\begin{align}
	\label{eq:ISC_def}
        \ISC&\equiv \ISC_1+\ISC_2 = \int_{n+1} \biggl( d\sigma_{\ISC_1} \Theta (\rcut - r^{\ISC_1}) +d\sigma_{\ISC_2} \Theta (\rcut - r^{\ISC_2})  \biggr) \,,
\end{align}
where $r^{\ISC_1}$ and $r^{\ISC_2}$ are approximations of the resolution variable in the limit where the radiated parton with momentum $k$ becomes collinear to the initial-state parton $p_1$ and $p_2$, respectively.
The differential cross section $d\sigma_{\ISC_i}$ includes the real phase space $d\Pi_R(\{p_j\},k)$, the real matrix element in the collinear limit $k \cdot p_i \to 0$ and the convolution with the PDFs.
In the following, we will provide a proper parametrisation for the real phase space $d\Pi_R(\{p_{j}\},k)$ in the initial-state collinear limit.

Without loss of generality, we focus on the collinear limit $k \cdot p_1 \to 0$. We start from the expression of the real-emission phase space in $d=4-2\epsilon$ dimensions
\begin{equation}\label{eq:realPS}
    d\Pi_R(\{p_j\},k) =[dk] \,
    \prod_{i=3}^{n+2}\,[dp_i] \,[dp_F] \,(2\pi)^d\delta^{(d)}(p_1+p_2-k-\sum_{i=3}^{n+2}p_i-p_F) \, ,
\end{equation}
where we have used the short hand notation $[dp]$ for the 1-particle phase space element
\begin{equation}
	[dp] \equiv \frac{d^d p}{(2\pi)^{d-1}}\delta^{+}(p^2-m^2) \,.
\end{equation}
We introduce the four-momentum $q^{\mu} = p_1^{\mu} +p_2^{\mu} -k^{\mu} = p_F^{\mu} + \sum_{i=3}^{n+2} p_i^{\mu}$ of all final-state particles but the radiated parton, and we rewrite $d\Pi_R(\{p_j\},k)$ as
\begin{align}\label{eq:psISR}
d\Pi_R(\{p_j\},k) &=  [dk] \,  d^dq \,\delta^{(d)}(p_1+p_2-k -q) \,
     \prod_{i=3}^{n+2}\,[dp_i] \, [dp_F] \,(2\pi)^d\delta^{(d)}\left(q-\sum_{i=3}^{n+2}p_i-p_F\right) \notag \\
    &=  d^dq \,  [dk]\,\delta^{(d)}(p_1+p_2-k -q)
     d\Pi^{d}_n(q;p_F,\{p_i\}) \, ,
\end{align}
where $d\Pi^{d}_n(q;p_F,\{p_i\})$ is the $d$-dimensional Lorentz-invariant phase space for a particle of four-momentum $q$ splitting into $n$ partons $\{p_i\}_{i=3,\dots,n+2}$ plus a colourless system $p_F$. It is worth mentioning that the four-momentum $q$ has a non-zero transverse component with respect to the direction of the colliding protons.

The radiation phase space $[dk]$ can be parametrised as
\begin{align}
    [dk] = \frac{1}{4 (2\pi)^{3-2\epsilon}}(k_t^{2})^{-\epsilon}dk^{2}_{t}\frac{d\cos{\theta}}{1-\cos^{2}{\theta}}
    \,d\Omega_{2-2\epsilon}\,,
\end{align}
where $\theta$ is the polar angle with respect to the beam axis in the partonic centre-of-mass (CM) frame, $k_t = k^0\sin{\theta}$ is the transverse momentum of the radiation and $d\Omega_{2-2\epsilon}$ spans the directions in the {$(2-2\epsilon)$-dimensional} transverse space. Performing a change of variables, we can write the radiation phase space as
\begin{align}
    [dk] &= \frac{1}{4}(k_t^{2})^{-\epsilon}\,dk^2_t\frac{dz}{\sqrt{(1-z)^2-4 z k_t^2/Q^{2}}}
    \,\frac{d\Omega_{2-2\epsilon}}{(2\pi)^{3-2\epsilon}} \,,
\end{align}
where we defined $Q = \sqrt{q^2}$ and the energy fraction $z = \frac{Q^2}{\hat{s}}$, at fixed $\hat{s} = (p_1 + p_2)^2$.\\
Since we are interested in the radiation collinear to $p_1$, we can approximate $k$ with $(1-z)p_1$ in the argument of the delta-function in Eq.~\eqref{eq:psISR}, dropping power suppressed contributions in $k_t$. The final expression for the real-emission phase space valid at leading power is
\begin{align}\label{eq:psISR2}
    d\Pi_R(\{p_j\},k)  = \frac{(4\pi^2)^{\epsilon}}{32\pi^3}\,
    d\Omega_{2-2\epsilon}\,
    \frac{dk^2_t}{(k_t)^{2\epsilon}}\, \frac{dz}{\sqrt{(1-z)^2-4 z k_t^2/Q^{2}}}
    \,
    d\Pi^{d}_n(q;p_F,\{p_i\}_{i=3}^{n+2})\,.
\end{align}
The matrix element squared for the real emission process $b_1+b_2\to b+\{a_i\}+F$ is denoted as $|{\cal M}_{b_1b_2;b \{a_i\}}|^2$, and in the collinear limit it assumes the well-known form
\begin{equation}\label{eq:realemm}
    |\mathcal{M}_{b_1b_2;b \{a_i\}}|^2 \approx \frac{8\pi\asbare\mu_0^{2\epsilon}}{z p_1\cdot k}\hat{P}^{s s'}_{a_1b _1}(z,\hat{k}_t;\epsilon)\mathcal{T}_{a_1b_2;\{a_i\}}^{s s'}.
  \end{equation}
In Eq.~\eqref{eq:realemm} the squared matrix element is implicitly assumed to be averaged (summed) over the colours and polarisations of the initial (final) state partons.
Unless stated otherwise, we shall use the same convention for all the squared matrix elements appearing in the paper.
The $\epsilon$ dependence of the matrix elements is always understood.
For a given Born matrix element
   \begin{align}
	\M_{\A}^{c_1, c_2, \ldots; s_1, s_2, \ldots}\left(p_1, p_2, \ldots\right) \,,
\end{align}
where $\{c_1,c_2,\ldots\}$ and $\{s_1,s_2,\ldots\}$ denote colour and spin indices respectively,
we defined the spin polarisation tensor
\begin{align}
  &\mathcal{T}_{\A}^{s_\alpha s_\alpha^{\prime}}\left(p_1,\ldots,p_\alpha,\ldots\right) \notag \\
  & \equiv  \frac{1}{\mathcal{S}}\!\!\!\underset{\text {spins } \neq s_\alpha, s_\alpha^{\prime}}{\overline{\sum}} \underset{~\text {colours }}{\overline{\sum}} \mathcal{M}_{\A}^{c_1, c_2, \ldots; s_1,\ldots s_\alpha \ldots}\left(p_1, p_2, \ldots\right)
\left[\mathcal{M}_{\A}^{c_1, c_2, \ldots; s_1, \ldots s_\alpha^{\prime}  \ldots }\left(p_1, p_2, \ldots\right)\right]^{\dagger}\! , \label{eq:polTensor}
\end{align}
where $\overline{\sum}$ indicates an average (sum) over the spins and colours of initial (final) state partons. The spin of the parton with momentum $p_\alpha$ is not summed over. However, we include a factor $1/\mathcal{S}$ , where $\mathcal{S}$ corresponds to the number of polarisations of the parton $\alpha$ if it is an initial-state parton and $\mathcal{S}=1$ otherwise. The fermion spin indices are $s_{\alpha}=\pm1$ while it is convenient to label the gluon spin $s_{\alpha}$ with the corresponding Lorentz index $\mu=1,\dots,d$. 
In Eq.~\eqref{eq:realemm} $\hat{P}^{s s'}_{a_1 b_1}$ is the unregularised Altarelli-Parisi splitting function for a splitting $b_1(p_1) \to a_1(zp_1) + b((1-z)p_1)$ defined in Appendix \ref{sec:APkernels}. The spin indices in the polarisation tensor defined in Eq. (\ref{eq:polTensor}) are those of the parton $a_1$ (i.e. the one that undergoes the collinear splitting).

We notice that the change of variable $\cos\theta \rightarrow z$ is not invertible for $\cos\theta \in [-1,1]$, and, therefore, the integrand has to be evaluated separately for positive (forward) and negative (backward) values of $\cos\theta$.
The radiation phase space element is the same in the two $\theta$-integration regions since it is an even function of $\cos\theta$. Furthermore, in the collinear limit, we can always choose an approximation of the resolution variable that is forward-backward symmetric.
Thus, the only contribution sensitive to the forward/backward direction is the collinear matrix element, and, more precisely, such a dependence is entirely due to the term $\frac{1}{p_1 \cdot k}$. Therefore, we can replace the latter term by the symmetric combination
\begin{equation} \frac{1}{p_1^0k^0}\biggl(\frac{1}{1-\cos\theta}+\frac{1}{1+\cos\theta}\biggr)=\frac{2(1-z)}{k_t^2},
\end{equation}
and consider only the integral in the interval $\cos\theta \in [0,1]$.
By combining the phase space parametrisation and the collinear approximation of the real matrix element for both initial-state collinear regions and summing over the Born channels, we derive the leading power contribution due to the initial-state collinear splittings as
\begin{align}	
	\label{eq:ISC_main}
	\ISC =  \sum_{\A} \sum_{\{b_1,b_2\}} \int_0^1 &dx_1 \int_0^1 dx_2\int  \frac{d\Pi^{d}_n(q ;p_F,\{p_i\}_{i=3}^{n+2})}{2Q^2} \notag \\
	&\times \int_{x_1}^1 \frac{dz_1} {z_1} \int_{x_2}^1 \frac{dz_2} {z_2}  \,
	\mathcal{I}_{a_1a_2}^{b_1b_2}(z_1,z_2)  f_{b_1}(x_1/z_1,\mu_F)  f_{b_2}(x_2/z_2,\mu_F)\, ,
\end{align}
where
\begin{align}
	\label{eq:I_ISC_general}
	\mathcal{I}_{a_1a_2}^{b_1b_2}(z_1,z_2) &=  \mathcal{T}_{a_1b_2;\{a_i\}}^{s s'} \mathcal{B}^{ss'}_{a_1 b_{1}}(z_{1},\rcut) \delta(1-z_{2}) +  \mathcal{T}_{b_1a_2;\{a_i\}}^{s s'}\delta(1-z_{1})\mathcal{B}^{ss'}_{a_2b_{2}}(z_{2},\rcut)  
\end{align}
is written in terms of the {\it cumulant} NLO beam function
\begin{align}\label{eq:beam_cum}
 \mathcal{B}^{ss'}_{ab} (z,\rcut)= & \left(\frac{\mu_R^2}{Q^2}\right)^{\epsilon}\frac{e^{\gamma_E \epsilon}}{\Gamma(1-\epsilon)}\frac{\as}{\pi}\notag\\
    &\times\int_0^{\infty} \!\!\frac{dx^{2}}{(x^2)^{1+\epsilon}} \int \frac{d\Omega_{d-2}}{\Omega_{d-2}}
    \, \hat{P}^{s s'}_{ab}(z,\hat{k}_t;\epsilon) \frac{\Theta(1-2x- z)}{\sqrt{1-{4x^2}/{(1-z)^{2}}}} \Theta (\rcut - r^{\ISC})
       \,.
\end{align}
In the above formul\ae\ we identified $Q^2 = x_1x_2 s$ and we introduced the dimensionless variable \mbox{$x^{2}= k_t^{2}/Q^{2}$}. In Eq.~\eqref{eq:beam_cum}, the upper limit in the integral over $z$, encoded in the theta function, comes from the small $x$ expansion of the physical solution of the equation \mbox{$1 -4zx^2/(1-z)^{2}\sim 1 -4x^2/(1-z)^{2}=0$}, associated with a vanishing argument of the square root appearing in the denominator, and it sets a kinematical endpoint $z<1$ for soft emissions.
We stress that, in a strict power counting in the pure collinear limit $x\to 0$, the square root factor appearing in Eq.~\eqref{eq:beam_cum} could be approximated with $1$ and, correspondingly, the upper limit in the $z$ integration extended to $1$. However, this procedure may lead to the appearance of rapidity divergences, depending on the specific observable under consideration. In particular, rapidity divergences appear for observables that behave as a transverse momentum in the collinear limit~\footnote{In SCET, the distinction is between SCET$_{\rm I}$, which do not need a rapidity regulator, and SCET$_{\rm II}$ observables, which do need it.}. We observe that retaining this term is consistent with a power counting valid both in the soft and collinear limits, according to the homogeneous scaling $x\sim \lambda$ and $1-z\sim\lambda$ for a small parameter $\lambda$.

We can further manipulate Eq.~\eqref{eq:beam_cum} under the assumption that, at fixed $k_{t}$, $r^{\ISC}$ is a regular function of $z$ in the soft limit $z\to 1$, which is valid for observables that scale as a transverse momentum in the initial-state collinear limit~\footnote{Notice that this is not the case for observables like $N$-jettiness that scales as $k_{t}^2/(Q^{2}(1-z))$.}. Then, for such observables we safely approximate
\begin{align}
\label{eq:isap}
 & \, \frac{\Theta(1-2x- z)}{\sqrt{1-4x^2/(1-z)^{2}}}
\hat{P}^{s s'}_{ab}(z,\hat{k}_t;\epsilon) = P^{s s'}_{ab}(z,\hat{k}_t;\epsilon)
+d_a^{s s'}\delta_{ab} \delta(1\!-\!z) \bigl(-\gamma_{a}\!-\!C_{a} \log(x) \bigr) + \mathcal{O}(x)\,,
\end{align}
where
\begin{align}
  \label{eq:dtensor}
d_a^{ss'} & = \begin{cases}
\delta^{ss'} &  a=q,{\bar q} \\
 -g^{\mu\nu}  &  a=g
 \end{cases}
\end{align}
and $P^{s s'}_{ab}$ are the regularised splitting kernels reported in Appendix \ref{sec:APkernels}.
We use
\begin{align}
 	C_a &= \begin{cases}
	C_F & ~a=q,{\bar q} \\
	C_A & ~a=g  
\end{cases}
\end{align}
and we define the coefficients
\begin{equation}
  \label{eq:gammas}
\gamma_q=\frac{3}{4}C_F\,,~~~~~~~~\gamma_g=\beta_0=\frac{11}{12}C_A -\frac{1}{3}T_Rn_f\, .
\end{equation}

\subsection{Final-state collinear limit}
\label{sec:FSC}
In this Section we outline the computation of the final-state collinear contribution in the region below the cut, $r<r_{\rm cut}$,
\begin{align}
	\label{eq:FSC_def}
    \FSC&\equiv  \sum_{i=3}^{n+2} \FSC_i =\sum_{i=3}^{n+2}  \int_{n+1} d\sigma_{\FSC_i} \Theta (\rcut - r^{\FSC_i})\, ,
\end{align}
where $r^{\FSC_i}$ refers to an approximation of the resolution variable in the limit where the radiated parton with momentum $k$ becomes collinear to the final-state parton $p_i$.
 In Eq.~(\ref{eq:FSC_def}), the differential cross section $d\sigma_{\FSC_i}$ includes the real phase space $d\Pi_R(\{p_j\},k)$, the real matrix element in the collinear limit $k \cdot p_i \to 0$ and the convolution with the PDFs.
For the sake of concreteness, we will focus on the region where $k$ becomes collinear to the final-state coloured parton with momentum $p_i$. In the following, we parallel the discussion carried out for initial-state radiation. We start from a suitable approximation of the real phase space in the relevant collinear limit. We consider Eq.~\eqref{eq:realPS} and we recast the phase space elements associated with the two collinear partons as
  \begin{equation}
    [dp_{i}][dk] = \frac{d^{d-1}{\vec p}_{i}}{(2\pi)^{d-1}2p_{i}^{0}} \frac{d^{d-1}{\vec k}}{(2\pi)^{d-1}2k^{0}} = \frac{d^{d-1}\vec{\tilde{p}}_{i}}{(2\pi)^{d-1}2\tilde{p}_{i}^{0}} \frac{\tilde{p}_{i}^{0}}{p_{i}^{0}}\frac{d^{d-1}{\vec k}}{(2\pi)^{d-1}2k^{0}} = [d\tilde{p}_{i}]\frac{\tilde{p}_{i}^{0}}{p_{i}^{0}}[dk]
\end{equation}
with $\tilde{p}_{i}^{\mu} = (\tp^{0}_{i},\vec{\tilde p}_{i}) = (p_{i}^{0}+k^{0},{\vec p}_{i} + {\vec k})$, from which it follows that the real phase space can be written as
\begin{equation}
  d\Pi_R(\{p_j\},k) =  d\Pi^{d}_n(q;p_F,\{p_3,\ldots,\tp_{i},\ldots,p_{n+2}\}) \frac{\tilde{p}_{i}^{0}}{p_{i}^{0}}[dk]\,,
\end{equation}
where $q = p_{1}+p_{2}$. We parametrise the radiation phase space in spherical coordinates as
\begin{equation}
 [dk] = \frac{1}{2(2\pi)^{3-2\epsilon}}(k^{0})^{1-2\epsilon} \sin^{1-2\epsilon}\theta dk^{0}d\theta d\Omega_{2-2\epsilon}\, ,
\end{equation}
where $\theta$ is the angle between ${\vec k}$ and $\vec{\tilde p}_i$. We reparametrise the phase space in terms of the energy fraction $\xi = k^{0}/\tp_{i}^{0}$ and the invariant mass $\tilde{s}_{i} = \tp^{2}_{i}$. Performing an expansion in small $\tilde{s}_{i}$ we obtain the following expression valid at leading power in the collinear limit
\begin{equation}
\frac{\tilde{p}_{i}^{0}}{p_{i}^{0}}[dk] =\frac{1}{4(2\pi)^{3-2\epsilon}} \,d\xi \xi^{-\epsilon}(1-\xi)^{-\epsilon} \,d\tilde{s}_i \tilde{s}_{i}^{-\epsilon}\left(1-\frac{\tilde{s}_{i}(1-2\xi)^{2}}{4(\tp_{i}^{0})^{2}\xi(1-\xi)}\right)^{-\epsilon} \,d\Omega_{2-2\epsilon}\, .
\end{equation}
This result agrees with standard collinear parametrisations, see for example Ref.~\cite{Mukherjee:2012uz}, apart from the last factor in parenthesis.
As discussed in the previous Section, this factor allows us to retain some terms which contribute beyond the strict collinear limit but make the integral finite without the need of introducing additional regulators. In particular, we count
 $\tilde{s}_{i}/\left[ (\tilde{p}_{i}^{0})^{2}\xi \right ] \sim 1$ ($\tilde{s}_{i}/\left[ (\tilde{p}_{i}^{0})^{2}(1-\xi) \right ] \sim 1$), where $\xi\to 0$ ($\xi\to 1$) corresponds to parton $k$ ($p_{i}$) becoming soft. Note that we can further approximate
\begin{equation}\label{eq:ps-factor-approx}
  \frac{\tilde{s}_{i}(1-2\xi)^{2}}{4(\tp_{i}^{0})^{2}\xi(1-\xi)} \sim \frac{\tilde{s}_{i}}{4(\tp_{i}^{0})^{2}\xi(1-\xi)}\,,
\end{equation}
which is valid in both limits $\xi\to 0$ and $\xi\to 1$.
The matrix element squared for the corresponding splitting process $a_i(\tp_i) \to a(k) + b(p_i)$ can be approximated in the collinear limit as
\begin{align}
    \label{eq:dsigma_FSC_definition}
    |\mathcal{M}_{a_1a_2;\dots a,b\dots}|^2 \approx \frac{8\pi\asbare\mu_0^{2\epsilon}}{k \cdot p_i}\hat{P}_{a_i\to ab}^{ss'}(\xi,\hat{k}_{\bot};\epsilon)\T_{a_1a_2;\dots a_i\dots}^{ss'}\, 
\end{align}
where $\hat{P}_{a_i\to ab}^{ss'}$ is the Altarelli-Parisi splitting function defined in Appendix \ref{sec:APkernels} and the spin-polarisation tensor $\T^{ss'}_{\A}$ is defined in Eq. (\ref{eq:polTensor}). In this case, the spin indices $s,s'$ in the polarisation tensor are those of the parton $a_i$ and $\hat{k}_{\bot}$ refers to the transverse momentum with respect to the direction of $\tilde{p}_{i}$.
The full  contribution $ \FSC_i$, associated with the collinear limit $k \cdot p_i \to 0$, is obtained by summing over all Born channels $\A=\{a_1,a_2,\{a_i\}\}$ and over all possible splittings $a_i\to (*)$ of the parton $a_i$. Thus, we find that
  \begin{align}
	\label{eq:xidefinition}
    \FSC_i &=  \sum_\A\int_0^1 \!\!dx_1 f_{a_1}(x_1,\mu_F)\!\int_0^1 \!\!dx_2 f_{a_2}(x_2,\mu_F)\!\int \!\!d\Pi^{d}_n(q;p_F,\{p_3,\ldots,\tp_{i},\ldots,p_{n+2}\}) \frac{\T_{a_1a_2;\dots a_i\dots}^{ss'}}{2Q^2} \notag\\
    &\times  \,\frac{\left(\frac{4\pi \mu_0^2}{Q^2}\right)^\epsilon}{\Gamma(1-\epsilon)}\frac{\asbare}{\pi} \,\sum_{(*)} \int \frac{d\Omega_{d-2}}{\Omega_{d-2}}\, \int_0^1 d\xi \xi^{-\epsilon}(1-\xi)^{-\epsilon} \notag\\
    &\times \,\int_0^{4(\tp_{i}^{0})^{2}/Q^2 \xi(1-\xi)} d \tilde{x}_{i}  \tilde{x}_{i}^{-1-\epsilon} \left(1-\frac{Q^{2}}{4(\tp_{i}^{0})^{2}}\frac{\tilde{x}_{i}}{\xi(1-\xi)}\right)^{-\epsilon} \hat{P}_{a_i\to (*)}^{ss'}(\xi,\hat{k}_\bot;\epsilon)\Theta(\rcut-r^{\FSC_i})\, ,
\end{align}
where we have introduced $Q = \sqrt{q^2}$ and $\tilde{x}_{i} = \tilde{s}_{i}/Q^2$. Notice that, at leading power, we can safely neglect the invariant mass $\tilde{s}_{i}$ in the phase space element $d\Pi^{d}_n$, which, thus, reduces to the Born-like phase space element for $n$ final-state massless partons and a colourless system $F$. With abuse of notation, we replace $\tp_i$ with $p_{i}$ everywhere so that the collection $(p_{F},\{p_{j}\}_{j=2}^{n+2})$ stands for a set of Born momenta.
Finally, the full final-state collinear contribution $ \FSC$ is obtained by summing over all collinear limits
\begin{align}
	\label{eq:FSC_main}
	\FSC=\sum_{\A}\sum_{i=3}^{n+2} \int_0^1 dx_1 f_{a_1}(x_1,\mu_F)\int_0^1 dx_2 f_{a_2}(x_2,\mu_F)\,\int  \frac{ d\Pi^{d}_n(q;p_F,\{p_{j}\}) }{2Q^2} \mathcal{I}_{a_1a_2;a_i}\, ,
\end{align}
where we introduced
\begin{align}
	\label{eq:I_FSC_general}
	\mathcal{I}_{a_1a_2; a_i} &=  \T_{a_1a_2;\dots a_i\dots}^{ss'} \mathcal{J}^{ss'}_{a_i}(\rcut)\, .
\end{align}
In the previous formula 
\begin{align}
	\label{eq:cum_jet_function}
	\mathcal{J}^{ss'}_{a_i}(\rcut) &= \Bigl(\frac{\mu_R^2}{Q^2}\Bigr)^{\epsilon}\frac{e^{\gamma_E \epsilon}}{\Gamma(1-\epsilon)}\frac{\as}{\pi}\,\sum_{(*)}\, \int \frac{d\Omega_{d-2}}{\Omega_{d-2}}   \,\int_0^1 d\xi \,\xi^{-\epsilon}(1-\xi)^{-\epsilon}\notag\\
	& \hspace{-0.5cm} \times\,\int_0^{4(p_{i}^{0})^{2}/Q^2 \xi(1-\xi)} d x_{i}  x_{i}^{-1-\epsilon} \left(1-\frac{Q^{2}}{4(p_{i}^{0})^{2}}\frac{x_{i}}{\xi(1-\xi)}\right)^{-\epsilon}\hat{P}_{a_i \to (*)}^{ss'}(\xi,\hat{k}_\bot;\epsilon)\Theta(\rcut-r^{\FSC_{i}})
\end{align}
is the cumulant NLO jet function. 
In Appendix \ref{app:Scet-like-jet-function} we outline a different approach, based on the method of regions, to define jet functions.

\subsection{Soft limit}
\label{sec:SOFT}
The last singular region we need to consider is the soft one. We observe that our construction leads unavoidably to overlaps among the different soft and collinear approximations. We take care of removing any double counting in the soft contribution $\SOFT$  by defining a ``subtracted'' soft current $\textbf{J}_{\sub}^{2}$ from which the soft limits of all initial- and final-state collinear approximations have been subtracted. A similar strategy has been used in Refs.~\cite{Catani:2014qha,Buonocore:2021akg,Catani:2023tby}. More precisely, we write the ensuing contribution below the cut as
\begin{equation}
  \label{eq:soft_main}
  \SOFT \equiv \frac{\as}{\pi}\sum_{\A}  \!\int_0^1 \!\!dx_1 f_{a_1}(x_1,\mu_F)\int_0^1\!\! dx_2 f_{a_2}(x_2,\mu_F) \int \!\frac{d\Pi^{d}_n(q;p_F,\{p_{j}\})}{2Q^2}  \bra{\mathcal{M}_{\A}^{(0)}} \mathbf{S} \ket{\mathcal{M}_{\A}^{(0)}}
\end{equation}
in terms of the NLO soft function
\begin{equation}
  \label{eq:S_general}
\mathbf{S} = 2\mu_R^{2\epsilon}  \,\frac{e^{\gamma_E \epsilon}}{\Gamma(1-\epsilon)}\,\int \frac{d^dk}{\Omega_{d-2}} \delta_{+}(k^2)\textbf{J}^2_{\sub} \, .
\end{equation}
In the above equation, the soft subtracted current is defined as
\begin{align}
  \label{eq:J2sub}
	\textbf{J}^2_{\sub} &= \biggl( -\BT_1\cdot \BT_2\omega_{12} 
	-\sum_{i}(\BT_1\cdot \BT_i\omega_{1i} + (1\leftrightarrow 2) )
	-\sum_{i > j}\BT_i\cdot \BT_j\omega_{ij}    \biggr) \Theta(\rcut - r_S) \notag \\
	&-\biggl(\BT_1^2\omega^1_2 \Theta(\rcut - r_{C_{1},S}) + (1\leftrightarrow 2) \biggr) - \sum_i\,\BT_i^2\omega_{ C_i,S}\Theta(\rcut - r_{C_i,S})\, ,
\end{align}
in terms of the eikonal kernels $\omega_{\alpha \beta}$ and $\omega^{\alpha}_{\beta}$ given by
\begin{equation}\label{eq:soft-kernels-IS}
	\omega_{\alpha \beta} \equiv \frac{p_{\alpha} \cdot p_{\beta}}{(k\cdot p_{\alpha})(k\cdot p_{\beta})} = \omega^{\alpha}_{\beta} + \omega^{\beta}_{\alpha} ~,~~~~~ \omega^{\alpha}_{\beta} \equiv \frac{p_{\alpha} \cdot p_{\beta}}{(k\cdot p_{\alpha})(k\cdot (p_{\alpha} + p_{\beta}))} ~~~~~\forall \alpha,\beta=1,...,n+2
\end{equation}
and of the soft limit of the final-state splitting kernels $\omega_{C_i,S}$ given by
\begin{equation}\label{eq:soft-kernels-FS}
	\omega_{C_i,S} \equiv  \frac{p_i \cdot (p_1 + p_2)}{(k\cdot p_i)(k\cdot (p_1 + p_2))}  ~~~~~\forall i=3,...,n+2\, .
  \end{equation}
In Eq.~\eqref{eq:J2sub}, $r_S$ is the soft limit of the slicing variable, whereas $r_{C_{\alpha},S}$ refers to the soft limit of the respective $\alpha$-collinear approximation of $r$.

We observe that the first line in Eq.~\eqref{eq:J2sub} corresponds to the standard eikonal contribution \mbox{${\textbf{J}^2=-\sum \BT_\alpha\cdot \BT_\beta \omega_{\alpha\beta}}$}, while the second line corresponds to the collinear singular contributions that are explicitly subtracted in order to obtain the purely soft wide-angle remainder. The resulting soft function is a well-defined quantity in $d$ dimensions. This is to be contrasted with soft functions defined in SCET, which may require the introduction of suitable rapidity regulators, see e.g. the calculation of one-loop soft functions for $N$-jet processes at hadron colliders discussed in Ref.~\cite{Bertolini:2017efs}.
Note that the kernels $\omega^{\alpha}_{\beta}$ and $\omega_{C_i,S}$, corresponding to the soft limit of the Altarelli-Parisi splitting functions, take the explicit form given in Eq.~\eqref{eq:soft-kernels-IS} and ~\eqref{eq:soft-kernels-FS} as a consequence of our use
of the energy fractions $z$ and $\xi$ in the initial-state and final-state collinear regions, respectively.

\subsection{Virtual contribution}

The virtual diagrams always contribute to the region below the slicing cut, $\rcut$, since, by definition, a proper slicing variable vanishes on a Born-like kinematic configuration. 
The $\overline{\text{MS}}$ renormalised on-shell scattering amplitude $\ket{ \M_\A(\mu_R^2, \{p_{\alpha}\})}$ can be perturbatively expanded as\footnote{The overall dependence on $\as$ entering at Born level is understood.}
\begin{equation}
	\ket{ \M_\A(\mu_R^2, \{p_{\alpha}\})} = \ket{ \M_\A^{(0)}(\{p_{\alpha}\})} + \frac{\as}{\pi}\ket{ \M_\A^{(1)}(\mu_R^2, \{p_{\alpha}\})} +  \mathcal{O}(\alpha^2_{\rm S})\, .
\end{equation}
It is related to the IR-finite amplitude $\ket{ \M_\A^{\text{fin}}(\mu_R^2, \{p_{\alpha}\})}$ via
\begin{equation}
	\ket{ \M_\A^{\text{fin}}(\mu_R^2, \{p_{\alpha}\})} = \bigl[1-\textbf{I}(\epsilon, \mu_R^2, \{p_{\alpha}\}) \bigr]\ket{ \M_\A(\mu_R^2, \{p_{\alpha}\})} \,,
\end{equation}
where $\textbf{I}(\epsilon, \mu_R^2, \{p_{\alpha}\})$ is the IR subtraction operator that admits the perturbative expansion
\begin{equation}
	\textbf{I}(\epsilon, \mu_R^2, \{p_{\alpha}\}) = \frac{\as}{\pi}\textbf{I}^{(1)}(\epsilon, \mu_R^2, \{p_{\alpha}\}) +  \mathcal{O}(\alpha^2_{\rm S})\,.
\end{equation}
In particular, we are interested in the one-loop finite remainder 
\begin{equation}
	\ket{ \M_\A^{(1),\text{fin}}(\mu_R^2, \{p_{\alpha}\})} = \ket{ \M_\A^{(1)}(\mu_R^2, \{p_{\alpha}\})} - \textbf{I}^{(1)}(\epsilon, \mu_R^2, \{p_{\alpha}\})\ket{ \M_\A^{(0)}(\{p_{\alpha}\})}\, ,
\end{equation}
where $\textbf{I}^{(1)}$ embodies the IR singular structure of the one-loop amplitude \cite{Giele:1991vf,Kunszt:1994np,Catani:1996vz}.
The explicit expression of $\textbf{I}^{(1)}(\epsilon, \mu_R^2, \{p_{\alpha}\})$ is
\begin{align} 
	&\textbf{I}^{(1)}(\epsilon, \mu_R^2, \{p_{\alpha}\}) = \left( \frac{\mu_R^2}{Q^2} \right)^{\epsilon} \frac{e^{\gamma_E \epsilon}}{\Gamma(1-\epsilon)} \frac{1}{4}
	\biggl\{ \sum_{\alpha\in \A} \left(
	-\frac{C_{\alpha}}{\epsilon^2} -\frac{2}{\epsilon}\gamma_{\alpha} \right) -\frac{2}{\epsilon} i\pi \BT_1\cdot \BT_2  \notag \\
	&\!+  \frac{2}{\epsilon}\sum_{i} \BT_i \cdot \left[ \BT_1\log\!\left( \frac{Q^2}{2p_1\cdot p_i}\right)\!+\!\BT_2\log\!\left( \frac{Q^2}{2p_2\cdot p_i}\right)\right]
	\!+\!  \frac{1}{\epsilon}\sum_{i\ne j} \BT_i\cdot \BT_j \left[ \log\!\left( \frac{Q^2}{2p_i\cdot p_j} \right) -i\pi \right]\biggr\} \, ,
\end{align} 
where the coefficients $\gamma_q$ and $\gamma_g$ are defined in Eq.~(\ref{eq:gammas}). 
It is useful to introduce the hard function
\begin{equation}
H(\alpha_\mathrm{S}(Q)) = 1 + \frac{\alpha_\mathrm{S}(Q)}{\pi} H^{(1)} +  \mathcal{O}(\alpha^2_{\rm S})\, ,
\end{equation}
where the $\mathcal{O}(\alpha_\mathrm{S})$ contribution is 
\begin{equation}
  	H^{(1)} \equiv\frac{\langle{ \M_\A^{(0)}(\{p_{\alpha}\})}\ket{ \M_\A^{(1),\text{fin}}(\mu_R^2=Q^2, \{p_{\alpha}\})}+ {\rm c.c.} }{| \M_\A^{(0)}(\{p_{\alpha}\}) |^2}\,,
\end{equation}
which is evaluated in $d=4$ dimensions.
The one-loop contribution to the NLO cross section is 
\begin{align}
	\VIRT &\equiv \frac{\as}{\pi} \sum_{\A}  \int_0^1 dx_1 f_{a_1}(x_1,\mu_F)\int_0^1 dx_2 f_{a_2}(x_2,\mu_F)\nonumber\\
  &\times \int \frac{d\Pi^{d}_n(q;p_F,\{p_{j}\})}{2Q^2} \, \left( \langle{ \M_\A^{(0)}(\{p_{\alpha}\})}\ket{ \M_\A^{(1)}(\mu_R^2, \{p_{\alpha}\})}+ {\rm c.c.} \right) \,,
\end{align}
where  $\A=\{a_1,a_2,\{a_i\}\}$ and
\begin{align}
	\langle{ \M_\A^{(0)}(\{p_{\alpha}\})}\ket{ \M_\A^{(1)}(\mu_R^2, \{p_{\alpha}\})}+ {\rm c.c.} &= \biggl( \bra{ \M_\A^{(0)}(\{p_{\alpha}\})} \textbf{I}^{(1)}(\epsilon, \mu_R^2, \{p_{\alpha}\})  \ket{ \M_\A^{(0)}(\{p_{\alpha}\})}  \notag \\
	&+\langle{ \M_\A^{(0)}(\{p_{\alpha}\})} \ket{ \M_\A^{(1),\text{fin}}(\mu_R^2, \{p_{\alpha}\})} \biggr) + {\rm c.c.} \;\;.
\end{align}
The manifest $\epsilon$-poles in the one-loop contribution cancel against those generated by the integration of 
 the real contribution over the radiation phase space, which are explicitly contained in the beam, jet and soft functions.
 The remaining singularities are of pure initial-state collinear origin and are cancelled by the $\overline{\text{MS}}$ counterterm associated with the renormalisation of the PDFs, which reads
 \begin{align}
   \label{eq:factct}
 	\CTMSB &  \equiv \sum_{\A}\sum_{\{b_1,b_2\}} \int_0^1 dx_1 \int_0^1 dx_2 \,\int \frac{d\Pi^{d}_n(q;p_F,\{p_{j}\})}{2Q^2}  \,
    \biggl(\frac{\mu_R^2}{\mu_F^2}\biggr)^{\epsilon} \frac{e^{\gamma_E \epsilon}}{\Gamma(1-\epsilon)}\frac{\as}{\pi}\, \vert \mathcal{M}^{(0)}_{\A}\vert^2\,  \frac{1}{\epsilon}  \notag \\
    &\!\!\!\times\! \int_{x_1}^{1} \frac{dz_1}{z_1}\,\int_{x_2}^{1} \frac{dz_2}{z_2}
    \biggl(  P_{a_1 b_1}(z_1; \epsilon=0) \delta_{a_2b_2}\delta(1-z_2)  + (1 \leftrightarrow 2) \biggr)
     f_{b_1}\!\!\left(\frac{x_1}{z_1},\mu_F\right)  f_{b_2}\!\!\left(\frac{x_2}{z_2},\mu_F\right)  \, .
 \end{align}


 \section{Applications}
\label{sec:applications}
 
In this Section we apply the general formalism discussed in Sect.~\ref{sec:slicing} to two candidate variables for jet hadroproduction processes,
the one-jet resolution variable $\Delta E_t$ and the $n$-jet resolution variable $\ktness$.
We start by presenting explicit analytic results for both observables in Sect.~\ref{sec:DeltaET} and \ref{sec:ktness}, respectively.
Finally, in Sect.~\ref{sec:results} we present NLO numerical results for specific processes obtained using both slicing variables.

\subsection{$\DeltaET$ slicing}
\label{sec:DeltaET}

%
We start by considering a variable relevant for the class of processes in which a colourless system $F$ is produced in association with a hard jet of momentum $p_J$.
For these processes a possible resolution variable can be defined as follows. Considering an event in which $F$ is accompanied by $m$ QCD partons with momenta $p_3,\dots,p_{m+2}$, we define
\begin{equation}
    \DeltaET = \sum_{i=3}^{m+2} \,|\vec{p}_{i,t}| - |\vec{p}_{F,t}| \,.
\end{equation}
At LO we have $m=1$, implying that $\DeltaET = 0$ for the Born and one-loop contributions, while for real-emission diagrams we have $m=2$.
Momentum conservation and the triangle inequality imply that $\DeltaET$ is non-negative and it vanishes only when all $\vec{p}_{i,t}$ are
either zero or antiparallel to $\vec{p}_{F,t}$. As a consequence, at NLO, real-emission diagrams lead to $\DeltaET = 0$ in the soft and/or collinear limits.

Thus, we can define the dimensionless slicing variable 
\begin{align}
	r=\frac{\DeltaET}{Q}\,
\end{align}
where $Q^2=(p_F+p_J)^2$ is the squared invariant mass of the Born-like system \footnote{$p_J$ can be determined with an arbitrary exclusive (always giving us exactly one jet) IR-safe jet-clustering algorithm. The choice of the algorithm will only affect the power corrections in $\rcut$.}.
The variable $\Delta E_t$ defined above is not expected to feature non-global logarithmic contributions \cite{Dasgupta:2001sh} and can be evaluated without relying on a jet algorithm. 
Therefore, it can be potentially useful to deal with the class of processes in which a colourless system $F$ is produced in association with a hard jet.
We will show in the following that this variable presents a richer structure compared to the one associated with $q_{T}$ or $1$-jettiness.
In view of these interesting features, the necessary ingredients to construct a slicing method turn out to be more difficult to evaluate.
In particular, $\DeltaET$ features a non-trivial azimuthal dependence which is responsible for the presence of non-vanishing spin correlations, already at NLO, and of additional initial-state collinear contributions with respect to those appearing in $q_T$-subtraction.

\subsubsection{Initial-state collinear limit}
\label{sec:ISC_DeltaET}
In the limit in which the radiated parton of momentum $k$ is collinear to the beams ($k_t \to 0$), we can write the transverse momentum of the final-state hard parton $p_3$ as 
\begin{equation}
    |\vec{p}_{3,t}| = |\vec{p}_{F,t} + \vec{k}_{t}| = \sqrt{|\vec{p}_{F,t}|^2 + |\vec{k}_{t}|^2 + 2|\vec{p}_{F,t}||\vec{k}_{t}|\cos\phi} \approx |\vec{p}_{F,t}|\biggl( 1+\frac{|\vec{k}_{t}|}{|\vec{p}_{F,t}|}\cos\phi\biggr)
\end{equation}
by neglecting quadratic terms in $k_t$. From this approximation, the normalised slicing parameter is
\begin{equation}
\label{eq:DeltaET_ISC}
     r^{\ISC} = r^{\ISC_1} = r^{\ISC_2} \equiv  \frac{\DETISC}{Q} \approx \frac{|\vec{p}_{F,t}|}{Q}\biggl( 1+\frac{|\vec{k}_{t}|}{|\vec{p}_{F,t}|}\cos\phi\biggr) + |\vec{k}_{t}|- |\vec{p}_{F,t}|= \frac{k_t}{Q}(1+\cos \phi)\, .
\end{equation}
For the sake of comparison, we introduce a simple power counting in terms of the energy of the emitted parton $E$ and of the angle $\theta$ it forms with the relevant collinear direction. In this region, we notice that the variable scales as $\DETISC \sim E \theta$, which is the same scaling as $q_{T}$.

By exploiting the parametrisation outlined in Sect.~\ref{sec: ISC}, the initial-state collinear contribution
in Eq.~(\ref{eq:ISC_main}) can be evaluated.
The function $\mathcal{I}_{a_1a_2}^{b_1b_2}(z_1,z_2)$ in Eq.~(\ref{eq:I_ISC_general}) reduces to
\begin{align}
	\label{eq:I_ISC_DET_result}
 \mathcal{I}_{a_1a_2}^{b_1b_2}(z_1,z_2) &= \biggl(\frac{\mu_R^2}{Q^2}\biggr)^{\epsilon}\frac{e^{\gamma_E\epsilon}}{\Gamma(1-\epsilon)}\frac{\as}{\pi}\, \vert \mathcal{M}^{(0)}_{\A}\vert^2\notag\\
    &\times\biggl\{ \biggl[ 
    \biggl( \frac{1}{2\epsilon^2} - \log^2(2\rcut)  - \frac{\pi^2}{3} \biggr) C_{a_1}\delta_{a_1b_1} \delta_{a_2b_2}\delta(1-z_1)\delta(1-z_2)  \notag \\
    & + \biggl( -\frac{1}{\epsilon} +2\log(2 \rcut) \biggr) \biggl( -\gamma_{a_1} \delta_{a_1b_1}\delta(1-z_1) 
    + P_{a_1 b_1}(z_1; \epsilon=0) \biggr) \delta_{a_2b_2}\delta(1-z_2)   \notag \\
    &+ C_{a_1 b_1}(z_1) \delta_{a_2b_2}\delta(1-z_2)  + (1 \leftrightarrow 2) \biggr]  \notag \\
    &+ \biggl[  G_{a_1b_1}(z_1) \delta_{a_1g}\delta_{a_2b_2}  \delta(1-z_2) \,\biggl( \frac{ |\mathcal{M}_{\A}^{(0),\epsilon_1}|^2}{\vert \mathcal{M}^{(0)}_{\A}\vert^2}
     - \frac{|\mathcal{M}_{\A}^{(0),\epsilon_2}|^2 }{\vert \mathcal{M}^{(0)}_{\A}\vert^2} \biggr)  + (1 \leftrightarrow 2) \biggr] \biggr\} \,,
\end{align} 
where $C_{ab}(z)$, $G_{ab}(z)$ and the regularised Altarelli-Parisi splitting kernels $P_{ab}(z)$ are defined in Appendix \ref{sec:APkernels}.
The spin-polarised matrix element, $|\mathcal{M}^{(0),\epsilon_i}_{\A}|^2$, is given by
\begin{equation}
	|\mathcal{M}^{(0),\epsilon_i}_{\A}|^2 = \mathcal{T}_{\A}^{\mu \nu} \epsilon_{i,\mu}\epsilon_{i,\nu} \,,
\end{equation}
where the gluon $a_1$ is polarised along $\epsilon_i^{\mu}$, with $i\in\{1,2\}$.  Here, $\epsilon_1^{\mu}$ is the transverse polarisation of the gluon along the transverse momentum of the colourless system with respect to the beam (i.e. $p_{F,t}$) and $\epsilon_2^{\mu}$ is the transverse polarisation orthogonal to $\epsilon_1^{\mu}$. More details on the spin-polarised matrix elements  $|\mathcal{M}^{(0),\epsilon_i}_{\A}|^2$ are provided in Appendix \ref{sec:AzInt}. 

The corresponding cumulant beam function in Eq.~(\ref{eq:beam_cum}) reads
  \begin{align}
    \label{eq:beam_deltaet}
\mathcal{B}^{ss'}_{ab} (z,\rcut)&=\biggl(\frac{\mu_R^2}{Q^2}\biggr)^{\epsilon}\frac{e^{\gamma_E\epsilon}}{\Gamma(1-\epsilon)}\frac{\as}{\pi}
    \biggl\{ d_a^{ss'}\biggl[ 
    \biggl( \frac{1}{2\epsilon^2} - \log^2(2\rcut)  - \frac{\pi^2}{3} \biggr) C_{a}\delta_{ab} \delta(1-z)  \notag \\
    & + \biggl( -\frac{1}{\epsilon} +2\log(2 \rcut) \biggr) \biggl( -\gamma_{a} \delta_{ab}\delta(1-z) 
    + P_{a b}(z; \epsilon=0) \biggr)+ C_{ab}(z) \biggr]  \notag \\
    &+  G_{ab}(z) \delta_{ag} {\cal A}^{ss'} 
       \biggr\} \,,
  \end{align}
  where $d_a^{ss'}$ is given in Eq.~(\ref{eq:dtensor}) and the asymmetry tensor ${\cal A}^{ss'}$ is defined
  in Appendix \ref{sec:AzInt}.
We see that $\DeltaET$ features a non-trivial azimuthal dependence which is responsible for the presence of non-vanishing spin-correlations, driven by ${\cal A}^{ss'}$ (see Eq.~(\ref{eq:AmunuTmunu})), already at this perturbative order.
Such contributions are new with respect to those appearing in $q_T$-subtraction.
This feature is analogous to what happens for the variable considered in Ref.~\cite{Chien:2020hzh}.

\subsubsection{Final-state collinear limit}
\label{sec:FSC_DeltaET}
In the following, frame-dependent quantities are specified in the partonic CM frame.
By using the parametrisation outlined in Sect.~\ref{sec:FSC}, we find that the slicing variable can be approximated as
\begin{align}
  \label{eq:rFSC}
     r^{\FSC} =\frac{\DETFSC}{Q} = \frac{Q}{2 p_{J,t}}x_{3} \sin^2\varphi
\end{align}
in the final-state collinear limit, where the angle between $p_3$ and $k$ goes to $0$. Here, $\cos\varphi=\frac{\vec{p}_{1,\bot}\cdot \vec{k}_\bot}{p_{1,\bot} k_\bot}$, where  $\vec{v}_\bot$ is obtained by projecting the spatial part of a four-vector $v$ onto the transverse plane of $\vec{p}_J$, and $p_{J,t}$ is the transverse momentum of the jet $p_J$ with respect to the beam direction. We notice that in this region the variable scales as $\DETFSC \sim E\theta^{2}$, which is the same scaling as $N$-jettiness in any collinear limit. Thus, the variable scales differently in the initial- and final-state collinear regions.
Finally, the final-state collinear contribution in Eq.~(\ref{eq:FSC_main})
can be evaluated. The function $ \mathcal{I}_{a_1a_2;a}$ in Eq.~(\ref{eq:I_FSC_general}) is found to be
\begin{align}
\label{eq:I_FSC_DET_result}
      \mathcal{I}_{a_1a_2;a} 
    &= \biggl(\frac{\mu_R^2}{Q^2}\biggr)^{\epsilon}\frac{e^{\gamma_E\epsilon}}{\Gamma(1-\epsilon)} \frac{\as}{\pi} \,|\mathcal{M}^{(0)}_{\A}|^2
    \,\biggl\{ \frac{1}{6} \biggl( \frac{C_A}{2} -T_R n_f \biggr)\biggl( \frac{|\mathcal{M}_{\A}^{(0),\varepsilon_1}|^2}{|\mathcal{M}^{(0)}_{\A}|^2}
     - \frac{|\mathcal{M}_{\A}^{(0),\varepsilon_2}|^2}{|\mathcal{M}^{(0)}_{\A}|^2} \biggr)\delta_{ag} \notag \\
    &
    - \gamma_a \biggl( - \frac{1}{\epsilon} +\log(8 \rcut)  +\frac{1}{2}\log\frac{\pjperp^2}{Q^2} \biggr) +\chi_a  \notag \\
    & +C_a\biggl[ \frac{1}{2\epsilon^2} -\frac{1}{2} \log^2(8\rcut) -\frac{1}{2}\log\frac{4 \lpjetbar^2}{Q^2}\left( \frac{1}{\epsilon}  -2\log(8 \rcut) \right)
    -\frac{1}{2}\log\frac{\pjperp^2}{Q^2} \log(8 \rcut)  \notag \\
    &  +\frac{1}{8}\log^2\frac{\pjperp^2}{Q^2} -\frac{1}{4}\left(  \log \frac{\pjperp^2}{Q^2} - \log \frac{4\lpjetbar^2}{Q^2}\right)^2\,  \biggr] 
    \biggr\} \, ,
\end{align}
where $\lpjetbar = \frac{p_J \cdot (p_1+p_2)}{Q}$ is the jet energy and we defined the constants
\begin{align}
	\chi_q&=C_F\left(\frac{7}{4}-\frac{5}{12}\pi^2\right)~~,& \chi_g&= \biggl( \frac{67}{36}-\frac{5}{12}\pi^2 \biggr)C_A - \frac{5}{9} T_{R}n_f \,.
\end{align}
$|\mathcal{M}^{(0),\varepsilon_i}_{\A}|^2$ is the spin-polarised matrix element 
\begin{equation}
	|\mathcal{M}^{(0),\varepsilon_i}_{\A}|^2 = \mathcal{T}_{\A}^{\mu \nu} \varepsilon_{i,\mu}\varepsilon_{i,\nu} \,,
\end{equation}
where the gluon $a$ is polarised along $\varepsilon_i^{\mu}$ with $i\in\{1,2\}$. Here, $\varepsilon_1^{\mu}$ is the transverse polarisation of the gluon along the transverse momentum of the beam with respect to the jet (i.e. $p_{1,\perp}$) and $\varepsilon_2^{\mu}$ is the transverse polarisation orthogonal to $\varepsilon_1^{\mu}$.
More details on the spin polarised matrix elements $|\mathcal{M}^{(0),\varepsilon_i}_{\A}|^2$ are provided in Appendix \ref{sec:AzInt}.

The corresponding cumulant jet function in Eq.~(\ref{eq:cum_jet_function}) reads
\begin{align}
  \label{eq:cumjetdeltaet}
\mathcal{J}^{ss'}_{a}(\rcut)&= \biggl(\frac{\mu_R^2}{Q^2}\biggr)^{\epsilon}\frac{e^{\gamma_E\epsilon}}{\Gamma(1-\epsilon)} \frac{\as}{\pi}
\,\Biggl\{ \frac{1}{6} \biggl( \frac{C_A}{2} -T_R n_f \biggr){\cal A}^{ss'}
    \delta_{ag} \notag \\
    & +d_a^{ss'}\Bigg[- \gamma_a \biggl( - \frac{1}{\epsilon} +\log(8 \rcut)  +\frac{1}{2}\log\frac{\pjperp^2}{Q^2} \biggr) +\chi_a  \notag \\
    & +C_a\Biggl( \frac{1}{2\epsilon^2} -\frac{1}{2} \log^2(8\rcut) -\frac{1}{2}\log\frac{4 \lpjetbar^2}{Q^2}\left( \frac{1}{\epsilon}  -2\log(8 \rcut) \right)
    -\frac{1}{2}\log\frac{\pjperp^2}{Q^2} \log(8 \rcut)  \notag \\
    &  +\frac{1}{8}\log^2\frac{\pjperp^2}{Q^2} -\frac{1}{4}\left(  \log \frac{\pjperp^2}{Q^2} - \log \frac{4\lpjetbar^2}{Q^2}\right)^2\,  \Biggr)\Bigg] 
    \Biggr\} \, ,
  \end{align}
  where $d_a^{ss'}$ is given in Eq.~(\ref{eq:dtensor}) and the asymmetry tensor ${\cal A}^{ss'}$ is defined in Appendix~\ref{sec:AzInt}.
We see that the non-trivial azimuthal dependence of $\DeltaET$ is responsible for non-vanishing spin-correlations also in this contribution.
As for the case of the initial-state collinear limit, this feature is analogous to what happens for the variable considered in Ref.~\cite{Chien:2020hzh}.

\subsubsection{Soft limit}
\label{sec:SOFT_DeltaET}
In the soft limit, $\DeltaET$ assumes the same expression (and, therefore, the same scaling in the soft-collinear limit) as the one derived in the initial-state collinear region, i.e. $r_S=r_{C_1,S} =r_{C_2,S}=\frac{k_t}{Q}(1+\cos \phi)$. We also need to specify the soft limit of the approximation we used in the final-state collinear region, which is given by $r_{C_3,S}= \frac{k\cdot p_3}{p_{J,t}}\frac{\sin^2\varphi}{Q}$.

By exploiting the results of Sect.~\ref{sec:SOFT}, the soft-subtracted contribution below the slicing cut, $\rcut$, can be written as
\begin{align}
	\SOFT = - \frac{\as}{\pi}&\sum_{\A} \,\int_0^1 dx_1 f_{a_1}(x_1,\mu_F)\int_0^1 dx_2 f_{a_2}(x_2,\mu_F)\times \notag \\
	&\times \int  \frac{d\Pi^{d}_1(p_1 + p_2 ;p_F,p_J)}{2Q^2}
\bra{\mathcal{M}_{\A}^{(0)}} \left[ \BT_1\cdot \BT_3\,\mathcal{S}_{13} +\BT_2\cdot \BT_3\,\mathcal{S}_{23}\right] \ket{\mathcal{M}_{\A}^{(0)}} 
\end{align}
where $\A=\{a_1,a_2,a\}$ labels the different Born channels $a_1+a_2\to a+F$. The soft integral $\mathcal{S}_{13}$ is defined as 
\begin{align}
    \mathcal{S}_{13} = 2\mu_R^{2\epsilon} \frac{e^{\gamma_E \epsilon}}{\Gamma(1-\epsilon)}  \,\int \frac{d^dk}{\Omega_{d-2}} \,\delta_+(k^2)  &
    \biggl\{\left(\frac{p_1\cdot p_J}{k\cdot p_1 k\cdot p_J}-\frac{p_1\cdot p_2}{k\cdot p_1 k\cdot (p_1+p_2)}\right)\Theta\left( \rcut - r_S \right) \notag\\
    &-\frac{(p_1+p_2)\cdot p_J}{ k\cdot (p_1+p_2) k\cdot p_J}\Theta\left(  \rcut - r_{C_3,S}\right)\biggr\}\,,
\end{align}
and ${\cal S}_{23}={\cal S}_{13}(p_1\leftrightarrow p_2)$.
 
After performing the integration over the radiation phase space, we obtain
\begin{align}
    {\cal S}_{13}= \left(\frac{\mu_R^2}{Q^2}\right)^\epsilon \frac{e^{\gamma_E \epsilon}}{\Gamma(1-\epsilon)}&
    \,\biggl\{ \left( \frac{1}{\epsilon} -2\log(8\rcut) \right)\biggl( \eta_{J} -\frac{1}{2}\log \frac{\pjperp^2}{Q^2}  +\frac{1}{2}\log \frac{4\lpjetbar^2}{Q^2}\biggr)
    \notag \\
    &+\frac{\pi^2}{12} + 4\log 2 \left(\eta_{J} + \log 2 \right) +\frac{1}{4}\left(  \log \frac{\pjperp^2}{Q^2} - \log \frac{4\lpjetbar^2}{Q^2}\right)^2
    \biggr\} \,,
\end{align}
where the jet rapidity is
\begin{equation}
  \eta_{J} =  \frac{1}{2}\log\biggl(\frac{p_2\cdot p_J}{p_1\cdot p_J} \biggr)\,.
\end{equation}

\subsubsection{Subtraction coefficients for $\DeltaET$ slicing}
\label{sec:DeltaETSigmas}
By adding all contributions we computed in the previous Sections, we manage to cancel the IR singular poles and we can now extract the expression for the 
$\BSigma^{1k}$ functions introduced in Sect.~\ref{sec:below_cut_general}:
\begin{align}
	\BSigma^{12}_{\A b_1 b_2}(z_1,z_2) &=  \delta_{a_1b_1} \delta_{a_2b_2} \delta(1-z_1)\delta(1-z_2)
	 \biggl\{   - C_{a_1} - C_{a_2} -\frac{C_a}{2}  \biggr\} \mathbf{1} \\
%
	\BSigma^{11}_{\A b_1 b_2}(z_1,z_2) &= 2\biggl(  P_{a_1 b_1}(z_1; \epsilon=0) \delta_{a_2b_2}\delta(1-z_2)  + (1 \leftrightarrow 2) \biggr)\mathbf{1}  \notag \\
	&+ \delta_{a_1b_1} \delta_{a_2b_2} \delta(1-z_1)\delta(1-z_2) \biggl \{  -( 2 \gamma_{a_1} + 2 \gamma_{a_2} + \gamma_a) +  \frac{C_a}{2}\log\frac{\pjperp^2}{Q^2}  \notag \\
	& - 2(C_A-C_F)( \delta_{a_1 g} - \delta_{a_2 g}) \eta_J 
	- ( 2 C_{a_1} + 2 C_{a_2} + 3C_a) \log2 
	\biggr\} \mathbf{1}  \\
%
	\BSigma^{10}_{\A b_1 b_2}(z_1,z_2) &=  \biggl[  \log\left(\frac{Q^2}{\mu_F^2} \right) + 2\log2 \biggr] \biggl(  P_{a_1 b_1}(z_1; \epsilon=0) \delta_{a_2b_2}\delta(1-z_2)  + (1 \leftrightarrow 2) \biggr)\mathbf{1} \notag \\
	&  +  \biggl( C_{a_1 b_1}(z_1) \delta_{a_2b_2}\delta(1-z_2)  + (1 \leftrightarrow 2) \biggr) \mathbf{1}  \notag \\
	&  + \biggl(  G_{a_1b_1}(z_1)  \delta_{a_1g}\delta_{a_2b_2} \delta(1-z_2) \,\biggl( \frac{ |\mathcal{M}_{\A}^{(0),\epsilon_1}|^2}{\vert \mathcal{M}^{(0)}_{\A}\vert^2}
     - \frac{|\mathcal{M}_{\A}^{(0),\epsilon_2}|^2 }{\vert \mathcal{M}^{(0)}_{\A}\vert^2} \biggr) + (1 \leftrightarrow 2)\biggr) \mathbf{1}   \notag \\
        	& + \delta_{a_1b_1} \delta_{a_2b_2} \delta(1-z_1)\delta(1-z_2) \biggl \{ H^{(1)} -p_{B}\beta_0\log\left(\frac{Q^2}{\mu_R^2} \right)  \notag \\
       	& - \left( C_{a_1} + C_{a_2} +\frac{C_a}{2} \right) \log^2 2    \notag \\
	& + \left(\!\!-(2\gamma_{a_1}  + 2\gamma_{a_2} + 3\gamma_a)  + 2(C_F-C_A)( \delta_{a_1 g} - \delta_{a_2 g}) \eta_J  +\frac{3}{2}C_a\log\frac{\pjperp^2}{Q^2} \right) \!\log{2}  \notag \\
     	& + \frac{C_a}{8}\log^2\frac{\pjperp^2}{Q^2}  -  \frac{\gamma_a }{2}\log\frac{\pjperp^2}{Q^2}  +\chi_a  - \left( C_{a_1} + C_{a_2} -\frac{C_a}{4} \right) \frac{\pi^2}{3} \notag \\
    	& + \frac{1}{6} \biggl( \frac{C_A}{2} -T_R n_f \biggr) \delta_{ag}  \biggl( \frac{ |\mathcal{M}_{\A}^{(0),\varepsilon_1}|^2}{|\mathcal{M}^{(0)}_{\A}|^2}
     	- \frac{|\mathcal{M}_{\A}^{(0),\varepsilon_2}|^2}{|\mathcal{M}^{(0)}_{\A}|^2} \biggr)   
     	\biggr\} \mathbf{1} \, ,
\end{align}
where $p_{B}$ denotes the power of $\as$ that appears in the LO cross section.

We note that, for a final-state emitter $l$, the contribution to the double logarithm in $\rcut$ is proportional to the Casimir ($C_l$) associated with the respective leg and it corresponds to half of the contribution from an initial-state leg.
This can be traced back to the different scaling behaviour of the slicing variable in the soft collinear limits.
If the variable scales as $E^{a_l} \theta^{b_l}$ in the limit in which the single emission is soft and collinear to leg $l$, 
this singular limit contributes to the coefficient of the double logarithm with a factor $-\frac{C_l}{a_l b_l}$.
In the specific case of $\DeltaET$, it turns out that $a_l = b_l = 1$ for initial-state radiation and $a_l = 1, b_l = 2$ for final-state radiation.


\subsection{$\ktness$ slicing}
\label{sec:ktness}
We now consider a more general class of processes in which an arbitrary number $n$ of hard jets is produced, possibly in association with a colourless system $F$.
Prominent examples among such processes are di-jet and tri-jet production, or the production of a colourless system in association with one or more hard jets.
The $\ktness$ variable, introduced in Ref.~\cite{Buonocore:2022mle}, takes its name from the $k_T$-clustering algorithm \cite{Catani:1993hr,Ellis:1993tq} and it represents an effective transverse momentum, describing the limit in which the additional jet is unresolved. If the unresolved radiation is close to the colliding beams or the event has no jets at Born level, $\ktness$ reduces to the transverse momentum ($q_T$) of the hard system. On the other hand, if the unresolved radiation is emitted close to one of the final-state jets, $\ktness$ describes the relative transverse momentum of the radiation with respect to the jet direction.

As already mentioned, the definition of $\ktness$ is based on the exclusive $k_T$-clustering algorithm, which is applied until $n+1$ final-state jets remain. If we are performing an N$^k$LO computation, the role of $\ktness$ is to discriminate between the fully unresolved region ($\ktness = 0$) and the region where at least one additional parton is resolved ($\ktness >0$). In the latter region, the IR-singularity structure can be at most of N$^{k-1}$LO-type. If we limit ourselves to NLO, the full clustering algorithm is not necessary and the definition of $\ktness$ directly coincides with the minimum among the usual distances $d_{ij}={\rm min}(p_{i,t}\,,p_{j,t}) \Delta R_{ij}/D$ between two particles
and the particle-beam distances $d_{iB}=p_{i,t} $. Here $D$ is a parameter of order unity and $\Delta R_{ij}$ is the customary $ij$ distance in rapidity and azimuth.
For a complete discussion of the recursive definition in the general case, we refer the reader to Ref.~\cite{Buonocore:2022mle},
where more details about the features of $\ktness$ are also provided.
In particular, besides being global, $\ktness$ turns out to be very stable with respect to hadronisation and multiparton interactions.

We can now define the dimensionless slicing variable
\begin{equation}
	r = \frac{\ktness}{Q}
\end{equation}
where $Q$ is the invariant mass of the hard system consisting of $n$ jets plus the colourless system. We point out that $Q$ must be an IR-safe quantity, and can for instance be determined by running the same IR-safe exclusive clustering algorithm exploited in the definition of $\ktness$, until exactly $n$ jets remain. 

\subsubsection{Initial-state collinear limit}
\label{sec:ISC_ktness}
When the unresolved radiation is collinear to an initial-state parton, $\ktness$ reduces to the usual transverse momentum of the Born-like system with respect to the beam, and, thus, it scales as $\sim E\theta$. It follows that the normalised slicing parameter can be approximated as
\begin{equation}
\label{eq:ktness_ISC}
     r^{\ISC} = r^{\ISC_1} = r^{\ISC_2} \equiv \frac{k_t}{Q} \,.
\end{equation}
By exploiting the parametrisation outlined in Sect.~\ref{sec: ISC}, the initial-state collinear contribution
in Eq.~(\ref{eq:ISC_main}) can be evaluated.
The function $\mathcal{I}_{a_1a_2}^{b_1b_2}(z_1,z_2)$ in Eq.~(\ref{eq:I_ISC_general}) reduces to
\begin{align}
	\label{eq:I_ISC_kTness_result}
 \mathcal{I}_{a_1a_2}^{b_1b_2}(z_1,z_2) &= \biggl(\frac{\mu_R^2}{Q^2}\biggr)^{\epsilon}\frac{e^{\gamma_E\epsilon}}{\Gamma(1-\epsilon)}\frac{\as}{\pi}\, \vert \mathcal{M}^{(0)}_{\A}\vert^2\notag\\
   & \times \biggl\{ \biggl[ 
    \biggl( \frac{1}{2\epsilon^2} - \log^2(\rcut)  \biggr) C_{a_1}\delta_{a_1b_1} \delta_{a_2b_2}\delta(1-z_1)\delta(1-z_2)  \notag \\
    & + \biggl( -\frac{1}{\epsilon} +2\log(\rcut) \biggr) \biggl(- \gamma_{a_1} \delta_{a_1b_1}\delta(1-z_1) 
    + P_{a_1 b_1}(z_1; \epsilon=0) \biggr) \delta_{a_2b_2}\delta(1-z_2)   \notag \\
    &+ C_{a_1 b_1}(z_1) \delta_{a_2b_2}\delta(1-z_2)  + (1 \leftrightarrow 2) \biggr]  
     \biggr\} \,,
\end{align} 
where $C_{ab}(z)$ and the regularised Altarelli-Parisi splitting kernels $P_{ab}(z)$ are defined in Appendix \ref{sec:APkernels}.
The cumulant beam function in Eq.~(\ref{eq:beam_cum}) reads
  \begin{align}
    \label{eq:beamktness}
     \mathcal{B}^{ss'}_{ab} (z,\rcut)=&d_a^{ss^\prime}\biggl(\frac{\mu_R^2}{Q^2}\biggr)^{\epsilon}\frac{e^{\gamma_E\epsilon}}{\
\Gamma(1-\epsilon)}\frac{\as}{\pi}
    \biggl\{ \biggl[
    \biggl( \frac{1}{2\epsilon^2} - \log^2(\rcut)  \biggr) C_{a}\delta_{ab}\delta(1-z)  \notag \\
    & + \biggl( -\frac{1}{\epsilon} +2\log(\rcut) \biggr) \biggl(- \gamma_{a} \delta_{ab}\delta(1-z)
    + P_{ab}(z; \epsilon=0) \biggr)+ C_{ab}(z)\biggr]
     \biggr\}\, ,
  \end{align}
where the tensor $d_a^{ss'}$ is defined in Eq.~(\ref{eq:dtensor}).  
The result in Eq. (\ref{eq:beamktness}) corresponds to the well-known transverse-momentum beam function, which appears in the production of a colourless system \cite{Collins:1984kg}. 

\subsubsection{Final-state collinear limit}
\label{sec:FSC_ktness}
In the following, frame-dependent quantities are specified in the partonic CM frame.
By using the parametrisation outlined in Sect.~\ref{sec:FSC}, we find that the slicing variable can be approximated as
\begin{align}\label{eq:ktenssFSCapprox}
     \left(r^{\FSC_i}\right)^2 = {\rm min}(\xi^2, (1-\xi)^2)\frac{x_i}{\xi(1-\xi) D^{2}}
\end{align}
in the final-state collinear limit,
where the angle between $p_i$ and $k$ goes to $0$. Here,
$D$ is the parameter entering the definition of $\ktness$. We notice that in this limit the variable scales as $\sim E\theta$, as expected from its definition as an effective transverse momentum with respect to any collinear direction. Thus, $\ktness$ features a uniform scaling in all initial-state and final-state collinear regions.

Finally, the final-state collinear contribution in Eq.~(\ref{eq:FSC_main})
can be evaluated. The function $ \mathcal{I}_{a_1a_2;a_i}$ in Eq.~(\ref{eq:I_FSC_general}) is found to be
\begin{align}
\label{eq:I_FSC_kTness_result}
      \mathcal{I}_{a_1a_2;a_i} 
    &= \biggl(\frac{\mu_R^2}{Q^2}\biggr)^{\epsilon}\frac{e^{\gamma_E \epsilon}}{\Gamma(1-\epsilon)} \frac{\as}{\pi} \,|\mathcal{M}^{(0)}_{\A}|^2
    \,\biggl\{ 
    -\gamma_{a_i} \biggl( -\frac{1}{\epsilon} + 2 \log(D \rcut) +2\log(2) \biggr) +\kappa_{a_i}  \notag \\
    & +C_{a_i} \biggl[ \frac{1}{2\epsilon^2} -\frac{1}{2\epsilon}\log \frac{4 E_{J_i}^2}{Q^2} - \log^2(D \rcut)  + \log \frac{4 E_{J_i}^2}{Q^2} \log(D \rcut)  \biggr] 
    \biggr\} \, ,
\end{align}
where the coefficients $\gamma_{a_i}$ are defined in Eq.~(\ref{eq:gammas}), $E_{J_i}=\frac{p_i \cdot (p_1+p_2)}{Q}$ is the energy of the $i$-th jet and we introduced the constants
\begin{align}
	\kappa_q&=C_F\left(\frac{7}{4}-\frac{\pi^2}{4} \right)~~,& \kappa_g&= \biggl( \frac{131}{72}-\frac{\pi^2}{4} \biggr)C_A - \frac{17}{36}T_{R}n_f \,.
\end{align}
The corresponding cumulant jet function in Eq.~(\ref{eq:cum_jet_function}) reads
\begin{align}
  \label{eq:cumjetktness}
\mathcal{J}^{ss'}_{a_i}(\rcut) &=d_{a_i}^{ss'}\biggl(\frac{\mu_R^2}{Q^2}\biggr)^{\epsilon}\frac{e^{\gamma_E \epsilon}}{\Gamma(1-\epsilon)} \frac{\as}{\pi}
    \,\biggl\{
    -\gamma_{a_i} \biggl( -\frac{1}{\epsilon} + 2 \log(D \rcut) +2\log(2) \biggr) +\kappa_{a_i}  \notag \\
    & +C_{a_i} \biggl[ \frac{1}{2\epsilon^2} -\frac{1}{2\epsilon}\log \frac{4 E_{J_i}^2}{Q^2} - \log^2(D \rcut)  + \log \frac{4 E_{J_i}^2}{Q^2} \log(D \rcut)  \biggr]
    \biggr\} \, .
\end{align}
\subsubsection{Soft limit}
\label{sec:SOFT_ktness}
In the following, frame-dependent quantities are specified in the partonic CM frame. 
In the soft limit, $\ktness$ assumes the expression
\begin{equation}
    r_S = \frac{k_{T,S}^{\rm ness}}{Q} = \min(1,\{\Delta R_{i(k)}\}/D)\frac{k_t}{Q} \,,
\end{equation}
where $\Delta R_{i(k)}$ is the distance between the soft parton with momentum $k$ and the hard parton $i$, and $i = 3,\dots,n+2$ runs over the final-state Born jets. 
We also need to specify the soft limit of the approximation we used in the singular region where the radiation is collinear to an initial-state parton, i.e  
$r_{C_1,S} =r_{C_2,S}=r^{\ISC}$, and in the singular region where the radiation is collinear to the final-state parton $i$, i.e.
\begin{equation}\label{eq:ktness2FSCS}
  \left(r_{C_i,S}\right)^2 = \frac{(k_{T,C_iS}^{\rm ness})^2}{Q^2} = \frac{k^{0}}{p^{0}_{i}} \frac{2 p_{i}\cdot k}{Q^2 D^2}\,,
\end{equation}
where $p_i$ is the four-momentum of the Born-level jet $i$.

By exploiting the parametrisation outlined in Sect.~\ref{sec:SOFT}, the soft-subtracted contribution
in Eq.~(\ref{eq:soft_main}) can be written in terms of the soft function in Eq.~(\ref{eq:S_general}).
The soft subtracted current $\textbf{J}^2_{\sub}$ in Eq.~(\ref{eq:J2sub}) becomes
\begin{align}
	\textbf{J}^2_{\sub} &= \biggl( -\BT_1\cdot \BT_2\omega_{12} 
	-\sum_{i}(\BT_1\cdot \BT_i\omega_{1i} + (1\leftrightarrow 2) )
	-\sum_{i > j}\BT_i\cdot \BT_j\omega_{ij}    \biggr)  \Theta(\rcut - r_S)  \notag \\
	&-\biggl(\BT_1^2\omega^1_2 + (1\leftrightarrow 2) \biggr)\Theta(\rcut - r^{\ISC}) - \sum_i\,\BT_i^2\omega_{C_i,S}\Theta(\rcut - r_{C_i,S})\, .
\end{align}
An analytical closed form for $\mathbf{S}$ is hard to obtain in this case. However, we can analytically extract the $\epsilon$-poles and logarithms of $\rcut$. 
In order to achieve this goal, we reorganise the subtracted current as follows
\begin{equation}
    \mathbf{J}^2_{\rm sub} = \mathbf{J}^2_{\rm sing} + (\mathbf{J}^2_{\rm sub}-\mathbf{J}^2_{\rm sing}) \equiv \mathbf{J}^2_{\rm sing} + \mathbf{J}^2_{\rm fin}
\end{equation}
where $\mathbf{J}^2_{\rm sing}$ is still singular in the soft wide-angle limit, whereas 
\begin{equation}
  \mathbf{S}_{\rm fin} \equiv  2 \mu_R^{2\epsilon} \frac{e^{\gamma_E \epsilon}}{\Gamma(1-\epsilon)}  \int \frac{d^d k}{\Omega_{d-2}} \delta_{+}(k^2)\mathbf{J}^2_{\rm fin}
\end{equation}
is finite in $d=4$ dimensions and can be computed numerically.

The soft-singular term $\mathbf{J}^2_{\rm sing} $ can be defined as
\begin{align}
    \mathbf{J}^2_{\rm sing}  &= \sum_i\, \BT_1 \cdot \BT_i \biggl( (\omega^1_2 - \omega^1_i)\Theta(\rcut - k_t/Q)
    + (\omega_{C_i,S} - \omega^i_1)\Theta(D\rcut - k_{i\perp}/Q) \biggr) + (1 \leftrightarrow 2) \notag \\
    &+ \sum_{i\ne j}\, \BT_i \cdot \BT_j (\omega_{C_i,S} - \omega^i_j)\Theta(D\rcut - k_{i\perp}/Q)
    \label{eq:J_sing}
\end{align}
where the sum runs over the labels of the final-state partons and $k_{i\perp}$ is the transverse momentum of $k$ with respect to the $i$-jet direction, in the partonic CM frame, (see Appendix \ref{sec:bspace_integral} for more details).  Note that we split the eikonal terms according to $\omega_{\alpha \beta}=\omega^\alpha_{\beta}+\omega^\beta_{\alpha}$ in order to obtain terms which are separately free of collinear divergences.
The integral of the singular part 
\begin{equation}
	\mathbf{S}_{\rm sing} \equiv   2 \mu_R^{2\epsilon} \frac{e^{\gamma_E \epsilon}}{\Gamma(1-\epsilon)}  \int \frac{d^d k}{\Omega_{d-2}} \delta_{+}(k^2)\mathbf{J}^2_{\rm sing}
\end{equation}
can be related to integrals that are already known from $q_T$-resummation for heavy-quark production (see Appendix \ref{sec:bspace_integral}). The final result can be written as
 \begin{align}
	\mathbf{S}_{\rm sing}=& \left(\frac{\mu_R^2 }{ Q^2}\right)^{\epsilon} \frac{e^{\gamma_E \epsilon}}{\Gamma(1-\epsilon)} \frac{\rcut^{-2 \epsilon}}{\epsilon}\biggl\{\sum_{i}C_{a_i}\log\left(\frac{2E_{J_i}}{Q}\right)+\frac{1}{2}\sum_{\alpha \neq \beta}\log \left(\frac{2p_\alpha \cdot p_\beta}{Q^2}\right) \BT_\alpha \cdot \BT_\beta \biggr\} \notag	\\
	&-\frac{1}{2}\biggl\{	2\log(D)\left[\sum_{\alpha\neq\beta} \BT_\alpha \cdot \BT_\beta\log\left(\frac{2p_\alpha\cdot p_\beta}{Q^2}\right)+2\sum_i C_{a_i} \log\left(\frac{2E_{J_i}}{Q}\right)\right]\notag\\
	&+\sum_i\,\biggl[\BT_1 \cdot \BT_i\biggl(
	{\rm Li}_2\biggl(-\frac{2p_2\cdot p_i}{Q^2} \biggr) +{\rm Li}_2\biggl(-\frac{p_2\cdot p_i}{2E_{J_i}^2}\biggr) \biggr) + (1\leftrightarrow 2)\biggr]\notag \\
	&+ \sum_{i\ne j}\,\BT_i \cdot \BT_j
	{\rm Li}_2\biggl( -\frac{E_{J_j}^2}{2(p_i\cdot p_j)}\sin^2\theta_{ij} \biggr)
	\biggr\}\, ,
\end{align}
where we recall that 
\begin{align}
	\cos\theta_{ij} &= 1-\frac{p_i\cdot p_j}{ E_{J_i}E_{J_j}} \,,& E_{J_i}&=\frac{p_i\cdot (p_1+p_2)}{Q}\,. 
\end{align} 
We point out that the logarithmic dependence on the jet energies $E_{J_i}$ cancels against the respective terms in the jet functions.

\subsubsection{Subtraction coefficients for $\ktness$ slicing}

By adding all contributions we computed in the previous Sections, and including the factorisation counterterm in Eq.~(\ref{eq:factct}), the IR singular poles cancel out and we can extract the expression for the 
$\BSigma^{1k}$ functions introduced in Eq.~(\ref{eq:below_cut_coefs}):
\begin{align}
	\BSigma^{12}_{\A b_1 b_2}(z_1,z_2) &=  -\delta_{a_1b_1} \delta_{a_2b_2} \delta(1-z_1)\delta(1-z_2)
	\biggl(C_{a_1}+C_{a_2}+\sum_{a \in \{a_i\}} C_{a} \biggr)  \mathbf{1} \\
	%
	\BSigma^{11}_{\A b_1 b_2}(z_1,z_2) &= 2\biggl(  P_{a_1 b_1}(z_1; \epsilon=0) \delta_{a_2b_2}\delta(1-z_2)  + (1 \leftrightarrow 2) \biggr)\mathbf{1}  \notag \\
	&+ \delta_{a_1b_1} \delta_{a_2b_2} \delta(1-z_1)\delta(1-z_2) \biggl \{ -2\biggl(\gamma_{a_1}+\gamma_{a_2}+\sum_{a\in \{a_i\}}\gamma_a\biggr)\mathbf{1} \notag \\
	&-2\log(D)\sum_{a\in\{a_i\}} C_a\mathbf{1}-\sum_{\alpha \neq \beta}\log \left(\frac{2p_\alpha \cdot p_\beta}{Q^2}\right) \BT_\alpha \cdot \BT_\beta 	\biggr\}   \\
	%
	\BSigma^{10}_{\A b_1 b_2}(z_1,z_2) &=   \log\left(\frac{Q^2}{\mu_F^2} \right) \biggl(  P_{a_1 b_1}(z_1; \epsilon=0) \delta_{a_2b_2}\delta(1-z_2)  + (1 \leftrightarrow 2) \biggr)\mathbf{1} \notag \\
	&  +  \biggl( C_{a_1 b_1}(z_1) \delta_{a_2b_2}\delta(1-z_2)  + (1 \leftrightarrow 2) \biggr) \mathbf{1}  \notag \\
	& + \delta_{a_1b_1} \delta_{a_2b_2} \delta(1-z_1)\delta(1-z_2) \biggl \{ \biggl( H^{(1)} -p_{B}\beta_0\log\left(\frac{Q^2}{\mu_R^2} \right) \biggr)\mathbf{1}\notag \\
	& +\sum_{a\in \{a_i\}} \biggl(-C_a \log^2\!D - 2\gamma_a  \log (2D) +\kappa_a\biggr) \mathbf{1} \notag\\
	&	-\log(D)\sum_{\alpha\neq\beta} \BT_\alpha \cdot \BT_\beta\log\left(\frac{2p_\alpha\cdot p_\beta}{Q^2}\right)\notag\\
	&-\frac{1}{2}\sum_i\,\biggl[\BT_1 \cdot \BT_i\biggl(
	{\rm Li}_2\biggl(-\frac{2p_2\cdot p_i}{Q^2} \biggr) +{\rm Li}_2\biggl(-\frac{p_2\cdot p_i}{2E_{J_i}^2}\biggr) \biggr) + (1\leftrightarrow 2)\biggr]\notag \\
	&-\frac{1}{2}\sum_{i\ne j}\,\BT_i \cdot \BT_j
	{\rm Li}_2\biggl( -\frac{E_{J_j}^2}{2(p_i\cdot p_j)}\sin^2\theta_{ij} \biggr)+ \mathbf{S}_{\rm fin} \biggr \}\,,
\end{align}
where $p_{B}$ denotes the power of $\as$ that appears in the LO cross section.
Contrary to what happens for $\Delta E_t$, the coefficient of the double logarithm in $\rcut$ is the same for both initial-state and final-state emitters, because $\ktness$ scales as $\sim E \theta$ in all soft-collinear limits.

\subsection{Numerical results}
\label{sec:results}
In this Section we present some numerical results obtained using $\Delta E_t$ and $\ktness$ as slicing variables.
We start by considering Higgs ($H$) boson production through gluon fusion in association with a jet at the LHC with a CM energy of 13 TeV.
We use the \texttt{PDF4LHC15\_nnlo\_30} PDFs \cite{Butterworth:2015oua} with $\alpha_S(m_Z) = 0.118$ through the \textsc{Lhapdf} interface~\cite{Buckley:2014ana}.
We define jets via the anti-$k_T$ clustering algorithm \cite{Cacciari:2008gp} with $R=0.4$ and $p_T^{j} > 30$~GeV. 
The factorisation ($\mu_F$) and renormalisation ($\mu_R$) scales are set to the Higgs boson mass $m_H=125$~GeV.

We compute the corresponding cross section (in the infinite top-mass limit) using both $\DeltaET$ and $\ktness$ as resolution variables.
The $\ktness$ calculation is carried out within the \Matrix framework \cite{Grazzini:2017mhc}, using tree-level and one-loop amplitudes evaluated with \OpenLoops~\cite{Cascioli:2011va, Buccioni:2017yxi,Buccioni:2019sur}. The $\DeltaET$ calculation is implemented in a dedicated code, which uses amplitudes computed with \Recola~\cite{Actis:2016mpe,Denner:2017wsf,Denner:2016kdg}.
To compare the results obtained with a $\DeltaET$ cut against those obtained with $\ktness$, we define the minimum $\rcut$ on the
dimensionless variable $r=\DeltaET/\sqrt{m_H^2 + (p_T^j)^2}$  and $r=\ktness/\sqrt{m_H^2 + (p_T^j)^2}$ respectively
\footnote{In comparing resolution variables having different scalings in the soft-collinear limits, it is possible to assign an exponent $p_X\neq 1$ to the definition of the dimensionless variable $r_{X} = \left(X/Q\right)^{p_{X}}$ associated with the resolution variable $X$ \cite{Campbell:2022gdq}. In the case of $\DeltaET$ the choice of $p_X$ would be non-trivial as this observable scales differently in the initial- and final-state regions.}.
\begin{figure}[H]
\begin{center}
\includegraphics[width=0.4\textwidth]{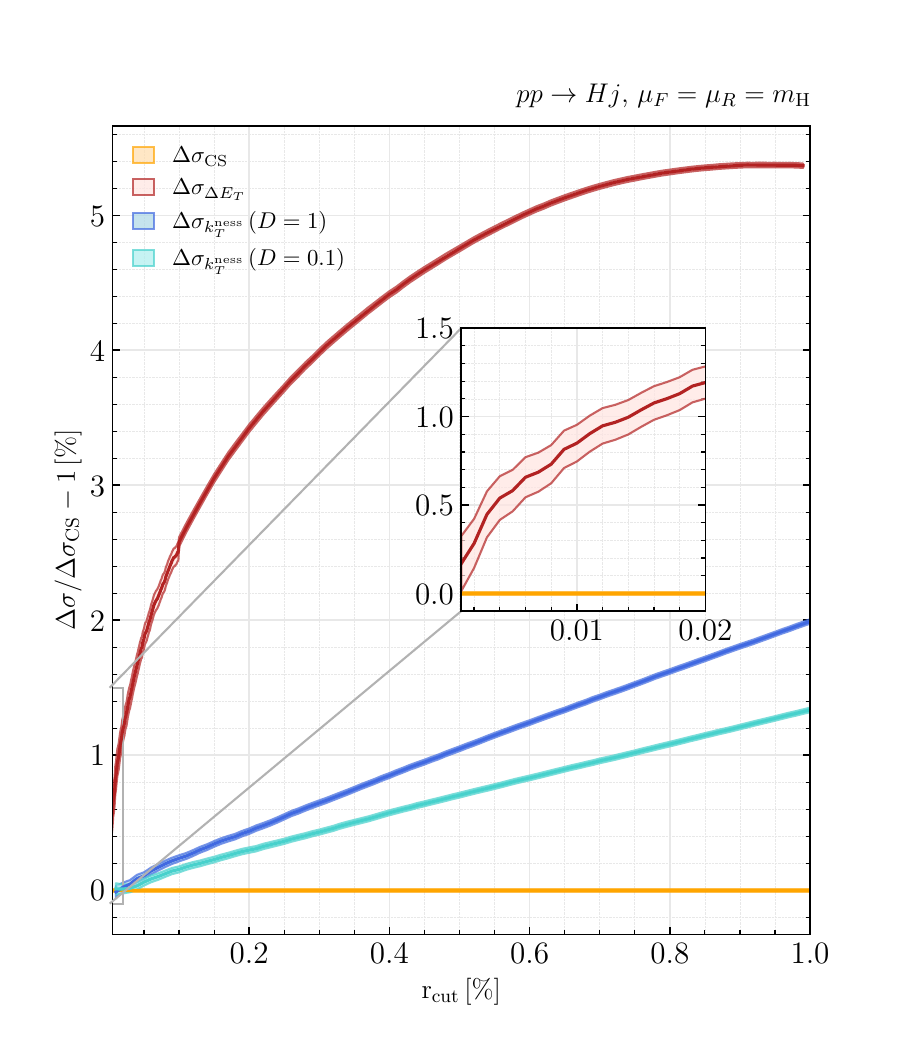}
\end{center}
\vspace{-0.5cm}
\caption[]{\label{fig:Hjet_unitarity}The NLO correction $\Delta\sigma$ to the $H+$~jet cross section computed with $\DeltaET$ (red curve)
and $\ktness$ (blue curves) as function of $\rcut$ compared to the $\rcut$-independent result obtained with CS subtraction using \Matrix (orange). In the case of $\ktness$, predictions for two different values of the parameter $D$, $D=1$ (dark-blue curve) and $D=0.1$ (light-blue curve), are shown.}
\end{figure}
In Fig.~\ref{fig:Hjet_unitarity} we study the behaviour of the NLO correction $\Delta\sigma$ as a function of $\rcut$ for the $\DeltaET$ and $\ktness$ calculations, normalised to the result obtained with Catani-Seymour (CS) dipole subtraction \cite{Catani:1996jh,Catani:1996vz} (which is independent of $\rcut$) by using \Matrix. Both calculations nicely converge to the expected result as $\rcut\to 0$, but the $\rcut$ dependence is very different for the two calculations.
The $\rcut$ dependence in the case of $\DeltaET$ is rather strong and consistent with a logarithmically enhanced linear behaviour.
By contrast, in the case of $\ktness$ the dependence is rather mild and linear for both values of $D$.
We observe that $\ktness$ slicing reaches $1\%$ accuracy at $\rcut=0.6\%$ ($\rcut=0.5\%$) for $D=0.1$ ($D=1$) while $\DeltaET$ slicing reaches the same accuracy at $\rcut=0.01\%$.

In Fig.~\ref{fig:Hjet_distributions} we show three relevant differential distributions, namely the invariant mass of the $H+1$~jet system, $m_{Hj}$ (left), the Higgs transverse momentum, $p_{T,H}$ (centre) and the rapidity of the leading jet, $y_{j1}$ (right). Each plot consists of three panels: in the upper panel we display the NLO differential cross section obtained with $\DeltaET$, $\ktness$ and CS subtraction; in the central panel we show the ratio between the NLO correction obtained with a slicing method ($\DeltaET$ in red, $\ktness$ in blue) and the NLO correction computed with CS subtraction (orange); in the lower panel we plot the NLO K-factor $K_{\rm NLO}$, defined as the ratio of the NLO to the LO distribution.
From the central panels we can observe a nice agreement between the results obtained with a slicing method and those computed with CS subtraction. For the $m_{Hj}$ and $p_{T,H}$ distributions the relative differences are around $1\%$ in the bulk region and below $2.5\%$ in the tails where the statistics is much lower and we still experience numerical fluctuations. Concerning the $y_{j1}$ distribution we observe an excellent control over the full rapidity range with differences smaller than $1\%$.

We also notice that the Higgs $p_T$ distribution displays a perturbative instability at $p_{T,H} = 30$ GeV which corresponds to the cut on the leading jet $p_T$. 
This behaviour is not physical but it is expected from a fixed-order computation: even if the observable is IR safe, an integrable divergence arises at a critical point inside the physical region where the distribution is not smooth. This divergence is associated with configurations where the Higgs boson is produced back-to-back to the leading jet in the transverse plane and the additional radiation can only be collinear to the beams or soft. The physical behaviour is restored when the all-order resummation of soft gluons is performed \cite{Catani:1997xc}.
\begin{figure}[H]
  \begin{center}
    \includegraphics[width=0.32\columnwidth]{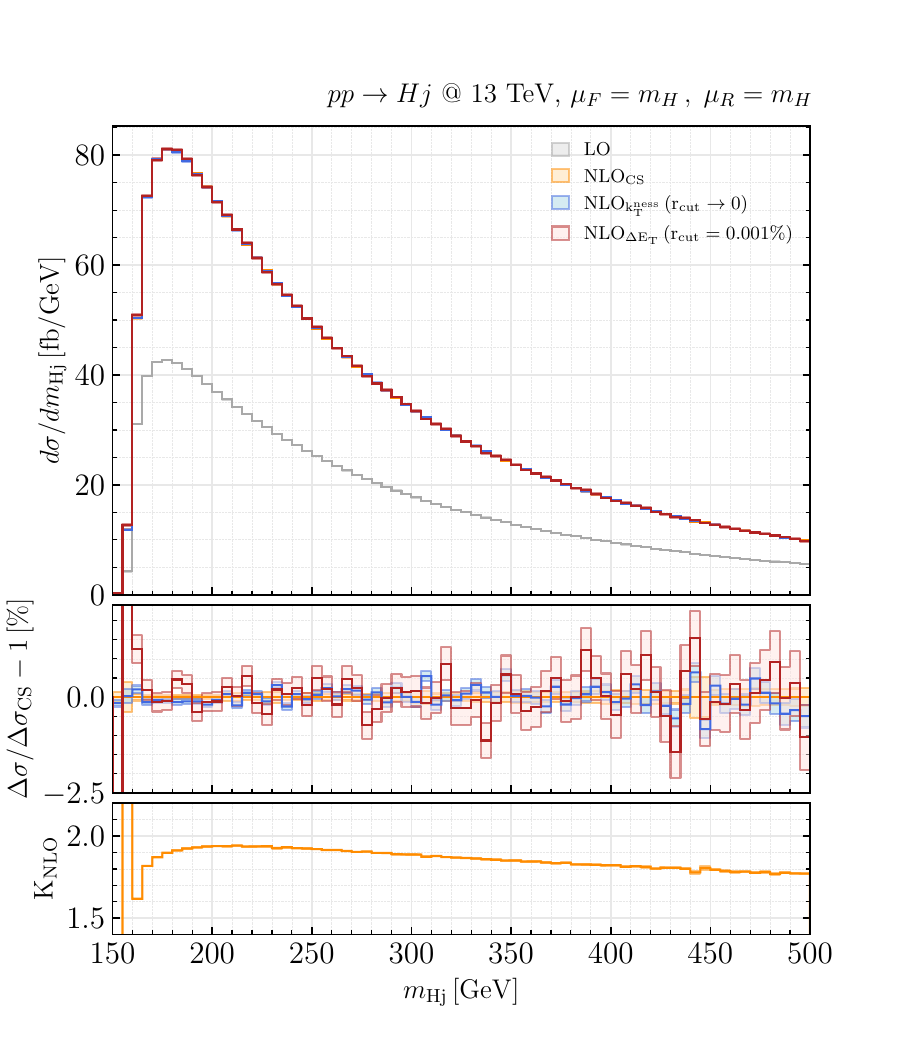}
    \hfill
    \includegraphics[width=0.32\columnwidth]{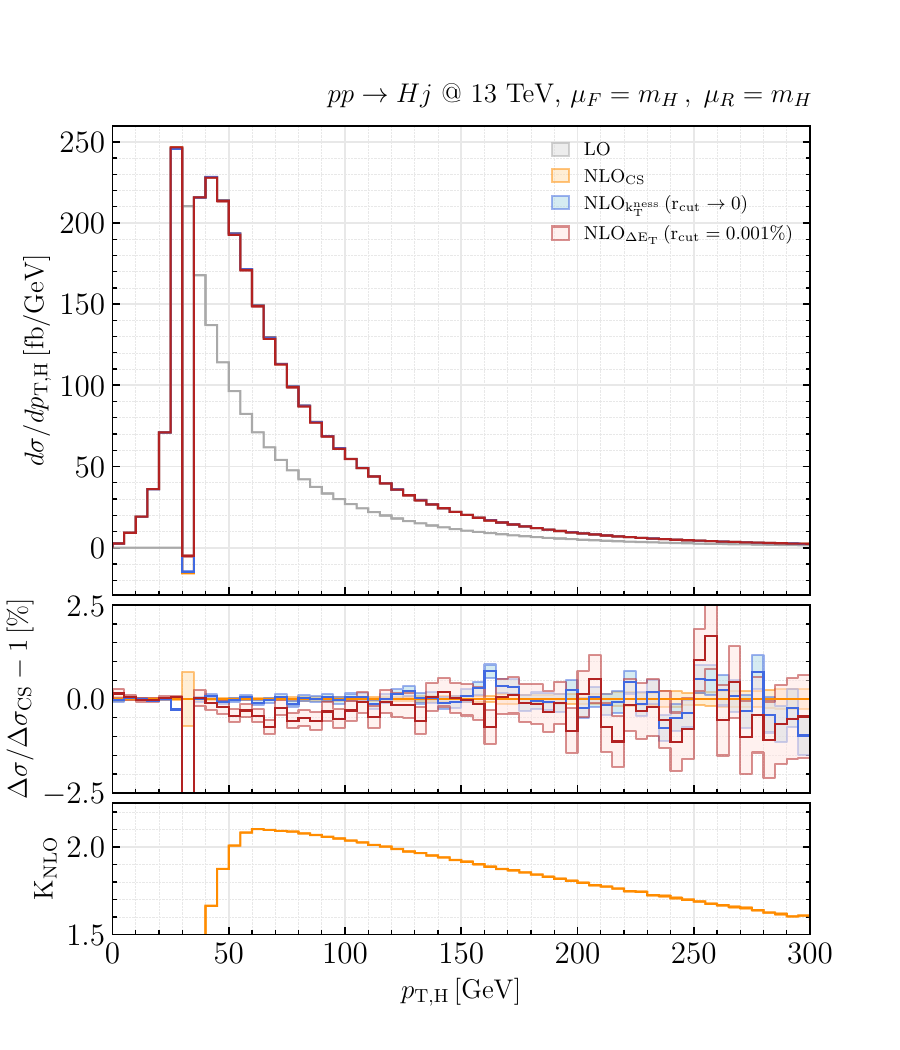}
    \hfill
    \includegraphics[width=0.32\columnwidth]{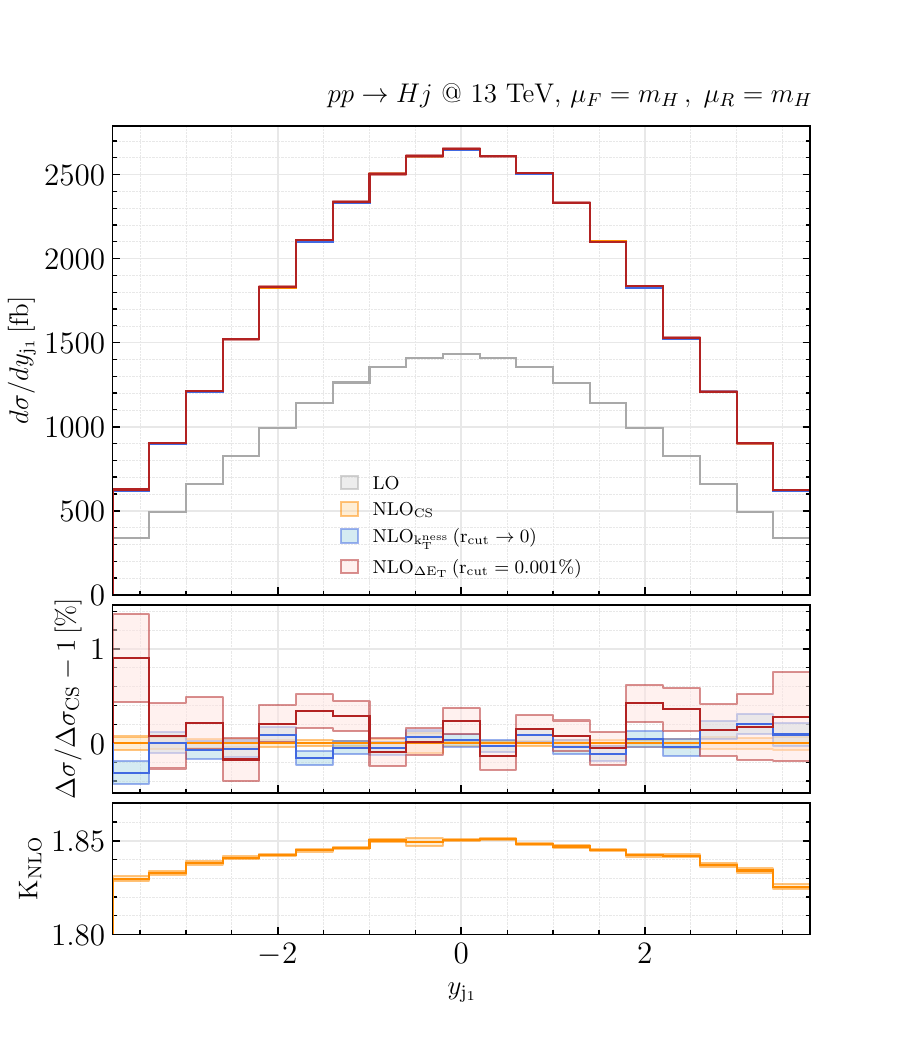}
    \hfill
\end{center}
\vspace{-0.5cm}
\caption{\label{fig:Hjet_distributions}
LO and NLO predictions for the invariant mass distribution of the $H+1$~jet system (left), the $p_T$ distribution of the Higgs boson (centre) and the rapidity distribution of the leading jet (right). The NLO correction is computed with $\DeltaET$, $\ktness$ and CS subtraction. In the case of $\DeltaET$, the results are obtained at a fixed $\rcut=0.001\%$, while, in the case of $\ktness$, they are obtained by performing a linear extrapolation to $\rcut \to 0$. For each distribution, absolute predictions are shown in the upper panel, ratios of the NLO correction computed with slicing approaches to the one computed with CS subtraction in the middle panel, and NLO K-factors in the lower panel.
}
\end{figure}

The $\ktness$ variable can be also used to evaluate multijet cross sections at NLO accuracy. Results for $Z+2$ jet processes were presented in Ref.~\cite{Buonocore:2022mle}.
Here we consider trijet production at the LHC with a CM energy of 13 TeV. 
We use the \texttt{NNPDF31\_nnlo\_as\_0118} PDFs \cite{NNPDF:2017mvq} and we require three jets in the final state with $p_T^{j} > 30$~GeV and $|\eta|^{j} < 4.5$.
The factorisation and renormalisation scales are set to the $Z$-boson mass $m_Z=91.1876$~GeV.
We compute the corresponding cross section using $\ktness$ as resolution variable (with $D=0.8$) and we define the minimum $\rcut$ on the
dimensionless variable $r=\ktness/m_{jjj}$ where $m_{jjj}$ is the invariant mass of the trijet system.

In Fig.~\ref{fig:jjj_unitarity} we study the behaviour of the NLO correction $\Delta\sigma$ as a function of $\rcut$ for the $\ktness$ calculation, normalised to the result obtained with CS dipole subtraction (which is independent of $\rcut$). We can clearly notice that the $\ktness$ slicing method nicely converges to the expected result and the $\rcut$ dependence is purely linear.
Compared to the case of $H+$jet production, the missing power corrections in $\rcut$ are much more pronounced (roughly a factor 30 larger) and at $\rcut=0.2\%$ we are still about $15\%$ away from the exact result.
We note however that, since the processes belong to different classes and involve different partonic channels at the Born level, it is not immediate to draw conclusions about the behaviour of the power corrections with respect to the number of jets.
Moreover, the hard scale used to normalise $\ktness$ in Fig.~\ref{fig:Hjet_unitarity} is not the same as the one in Fig.~\ref{fig:jjj_unitarity} and the NLO $K$-factor for $H$+jet is roughly a factor four larger than the one for trijet production.
If we had used the invariant mass of the Born-like system to define the dimensionless variable $r$ for $H$+jet, as we do in trijet production, and normalised the missing power corrections to the respective LO cross sections, the slope of the linear power corrections would only have differed by a factor four. In general, the quantitative impact of the linear power corrections may depend both on the hard scale appearing in the definition of the dimensionless variable $r$ and on the parameter $D$ in the $\ktness$ definition.
In the case of $H+$jet production we have seen in Fig.~\ref{fig:Hjet_unitarity} that reducing the parameter $D$ from $D=1$ to $D=0.1$ reduces the impact of power corrections in a significant way. In the case of trijet production we find instead that a reduction of $D$ from $D=0.8$ to $D=0.1$ leads to a marginal increase of the effect.
More detailed studies on these issues are left for future work.

\begin{figure}[H]
\begin{center}
\includegraphics[width=0.8\textwidth]{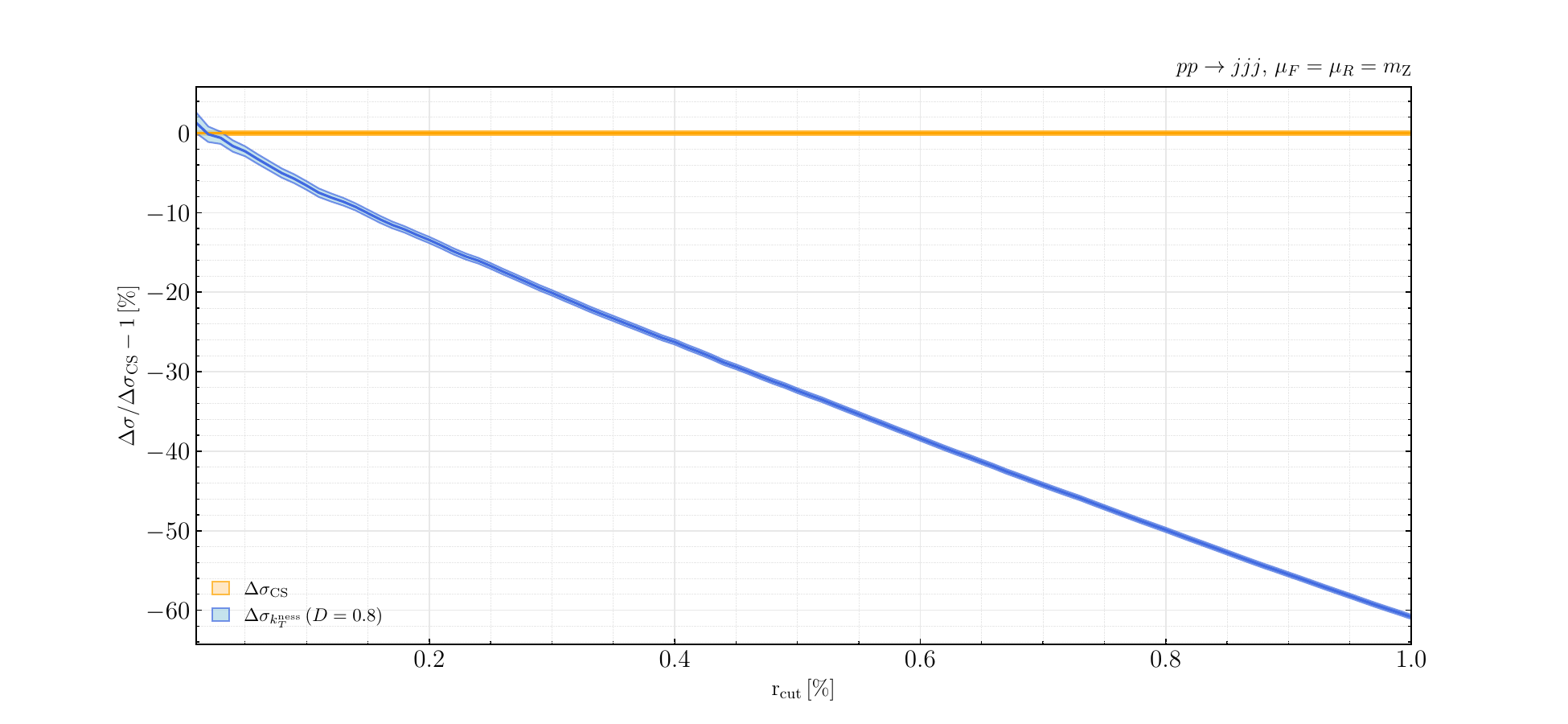}
\end{center}
\vspace{-0.5cm}
\caption[]{\label{fig:jjj_unitarity}
The NLO correction $\Delta\sigma$ to the trijet cross section computed with $\ktness$ (blue curve) as function of $\rcut$ compared to the $\rcut$-independent result obtained with dipole subtraction (orange curve).
}
\end{figure}

\section{Summary}
\label{sec:summa}

Slicing methods have provided very efficient ways to obtain higher order QCD predictions for a number of benchmark hadron collider
processes. These methods are based on identifying a resolution variable to distinguish configurations in which at least one additional QCD parton is resolved.
The exploration of new resolution variables can have interesting implications,
both in the context of fixed-order calculations, resummed computations and in Monte Carlo event generators.

In this paper we have considered a general class of hadron collider processes: the production of an
arbitrary number of jets, possibly accompanied by a colourless system. We have provided
a general formulation of a slicing scheme for this class of processes, by identifying the various
contributions that need to be computed at NLO.
Following the nomenclature customarily used in SCET,
these contributions are the parton level beam, jet and soft functions describing initial-state collinear, final-state collinear, and soft wide-angle radiation.

We then focused on two new observables, the
one-jet resolution variable $\DeltaET$ and the $n$-jet resolution variable $\ktness$ \cite{Buonocore:2022mle}, and we have explicitly computed
all the perturbative contributions needed to carry out NLO calculations by using these variables.
We have shown that the $\DeltaET$ variable, though potentially interesting for one-jet processes, has a non-trivial structure of
azimuthal correlations that lead to complications in the evaluation of the beam and jet functions already at NLO.
We have presented numerical results for $H$+jet production using $\DeltaET$ and $\ktness$, showing
the different power suppressed contributions affecting the two variables. While power corrections for $\ktness$ are purely linear,
in the case of $\DeltaET$ they are logarithmically enhanced. We have also shown results for differential distributions
obtained with the two slicing methods, and we have presented new results for three jet production obtained with $\ktness$.
Work on the extension of the $\ktness$ formalism to NNLO is ongoing, and will be reported elsewhere.

In a series of Appendices we provide extensive details of our calculations. We also
computed the jet function for $\DeltaET$ and $\ktness$ by using alternative SCET-like definitions,
which might be more suitable for an extension to NNLO.

\vskip 0.5cm
\noindent {\bf Acknowledgements}

\noindent This work is supported in part by the Swiss National Science Foundation (SNSF) under contracts 200020$\_$188464 and PZ00P2$\_$201878 and by the UZH Forschungskredit Grant FK-22-099. We would like to thank Stefano Catani for helpful discussions.
We are grateful to Stefan Kallweit for his support in the implementation of $\ktness$ in \Matrix and for comments on the manuscript.

\appendix
\appendixpage
\section{Azimuthal integrals for $\DeltaET$}
\label{sec:AzInt}

When we consider $\DeltaET$ as a resolution variable, both initial- and final-state collinear limits feature a non-trivial dependence on the azimuthal angle in the plane transverse to the collinear direction.
In this Appendix we will explain related complications that appear in the calculation of the $\I_{*}$-functions defined in Eqs. (\ref{eq:I_ISC_general}) and (\ref{eq:I_FSC_general}) for initial-state and final-state collinear limits, respectively.

In order to treat both collinear limits at the same time, we introduce the notation
\begin{align}
	\Phi \equiv \begin{cases}
	\phi \\
	\varphi
	\end{cases} &&
	x \equiv \begin{cases}
	z  \\
	\xi
	\end{cases} &&
	k_T \equiv \begin{cases}
	k_t  \\
	k_\bot 
	\end{cases}&& 
\P_{(*)}^{\mu \nu} \equiv \begin{cases}
		P_{(*)}^{\mu\nu}(z,\hat{k}_t;\epsilon) & \text{~~~~~~for ISC}\\
		\hat{P}_{(*)}^{\mu\nu}(\xi,\hat{k}_\bot;\epsilon) & \text{~~~~~~for FSC}
\end{cases}\,,
\end{align}
where
$(*)$ stands for an arbitrary splitting.
In the initial-state collinear limit (ISC) where a parton splits into a gluon entering the hard process, the corresponding splitting kernel contains a spin-dependent term $\kpmunu$ which is a function of the azimuthal angle $\Phi$. A similar situation occurs in the final-state collinear limit (FSC) when a gluon splits into two collinear partons.
If the resolution variable is independent of the azimuthal angle, one can straightforwardly perform the azimuthal integrals in Eqs. (\ref{eq:beam_cum}) and  (\ref{eq:cum_jet_function})    
\begin{equation}
	\int \frac{d\Omega_{d-2}}{\Omega_{d-2}}\P_{(*)}^{\mu\nu}(x,\hat{k}_T;\epsilon) = \P_{(*)}(x;\epsilon) (-g^{\mu\nu}) \,.
\end{equation}
This corresponds to replacing the spin-polarised splitting kernel $\P_{(*)}^{\mu\nu}(x,\hat{k}_T;\epsilon)$ with the averaged one $\P_{(*)}(x;\epsilon)$ in the approximation of the matrix element and using the fact that 
\begin{equation}
	-g_{\mu\nu} \T^{\mu\nu}_{\A} = |\mathcal{M}^{(0)}_{\A}|^2 \,,
\end{equation}
where $\A$ is the multi-index labelling a Born channel. 

In the general case, the splitting kernel can be decomposed as the sum of a spin-averaged part $ \P_{(*)}$  and a contribution proportional to $G_{(*)}$ that averages to zero when the slicing variable has a trivial dependence on the azimuthal angle, namely
\begin{equation}
	\P_{(*)}^{\mu\nu}(x,\hat{k}_T;\epsilon) = \P_{(*)}(x;\epsilon)(-g^{\mu\nu}) - G_{(*)}(x)\biggl( -g^{\mu\nu} - 2(1-\epsilon)\kpmunu  \biggr)
	\label{eq:APpolarised}\, ,
\end{equation}
where $\hat{k}^{\mu}_T = \frac{k^{\mu}_T}{|\vec{k}_T|} $ is the normalised transverse four-momentum.
All $\P_{(*)}$ and $G_{(*)}$ functions are listed in Appendix \ref{sec:APkernels}. 
If we plug Eq.~(\ref{eq:APpolarised}) in the expression of the $\I_{*}$ integrals in Eqs. (\ref{eq:I_ISC_general}) and (\ref{eq:I_FSC_general}), we end up with two contributions
for the spin-averaged and spin-dependent parts, respectively.
The azimuthal dependence of the slicing variable does not significantly increase the complexity of the
spin-averaged integral, thus we will focus on
the spin dependent part in the following.

We can first perform the integration over all radiation variables except the angular dependence in the transverse plane $d\Omega_{d-2}$. By dropping power corrections in the slicing parameter, $\rcut$, we arrive at integrals of type 
\begin{equation}
   \I_\Phi^{\mu\nu}  = \int \frac{d\Omega_{d-3}}{\Omega_{d-2}} \int_{0}^{\pi}d\Phi (\sin\Phi)^{-2\epsilon}\, g(\Phi) \left(-g^{\mu\nu} - 2(1-\epsilon)\kpmunu\right)
\end{equation}
to be contracted with the spin-polarised tensor $\T^{\mu\nu}_{\A} $.
The function $g(\Phi)$ embodies the remaining dependence on $\Phi$ after the integration over the transverse momentum: in the initial-state collinear limit we find $g(\phi)=(1+\cos\phi)^{2\epsilon}$, while in the final-state collinear limit we have $g(\varphi)=(\sin\varphi)^{2\epsilon}$.
In order to perform the integral of $\kpmunu$, we remind the reader that $\cos\phi=\frac{\vec{p}_{F,t}\cdot \vec{k}_t}{p_{F,t} k_t}$ for the initial-state collinear limit and $\cos\varphi=\frac{\vec{p}_{1,\bot}\cdot \vec{k}_{\bot}}{p_{1,\bot} k_\bot}$ for the final-state collinear limit.
Then, we can write the transverse unit vector as
\begin{align}
    \hat{k}_T^\mu= \cos\Phi\,  \E_1^\mu +\sin\Phi\, \E_2^\mu(\Omega_{d-3}) \,,
\end{align}
where we defined 
\begin{equation}
\E_1^\mu = \begin{cases}
	\epsilon_1^{\mu} = \frac{p_{F,t}^\mu }{|\vec{p}_{F,t}|}  &\text{for ISC} \\
	\varepsilon_1^{\mu} = \frac{p_{1,\bot}^\mu }{|\vec{p}_{1,\bot}|} &\text{for FSC}\,\,\,,
	\end{cases} 
\end{equation}
while $\E_2^\mu(\Omega_{d-3})$ lives on the unit sphere of the $(d-3)$-dimensional space orthogonal to $\E_1^{\mu}$.
The four-vectors $\E_1^{\mu}$ and $\E_2^{\mu}$ satisfy the conditions $\E_1^2=\E_2^2=-1$ and $\E_1\cdot \E_2=0$. 
It follows that
\begin{align}
    - 2(1-\epsilon) \int \frac{d\Omega_{d-2}}{\Omega_{d-2}} &g(\Phi) \,\kpmunu  = -\frac{2(1-\epsilon)}{\Omega_{d-2}}\int_0^\pi d \Phi (\sin\Phi)^{-2\epsilon}\,g(\Phi)\int d\Omega_{d-3}
\notag \\
&\times\left(\cos^2\Phi {\E_1}^\mu{\E_1}^\nu +\sin\Phi\cos\Phi \left({\E_1}^\mu{\E_2}^\nu+{\E_2}^\mu{\E_1}^\nu \right)+\sin^2\Phi {\E_2}^\mu{\E_2}^\nu \right) \,.
\end{align}
The integral of $\sin\Phi\cos\Phi \left({\E_1}^\mu{\E_2}^\nu\!\!+\!\!{\E_2}^\mu{\E_1}^\nu \right)$ vanishes because the integrand is anti-symmetric in the $(d-3)$-dimensional subspace.
The only non-trivial contribution to the $d\Omega_{d-3}$-integral is
\begin{align}
    \int d\Omega_{d-3}{\E_2}^\mu{\E_2}^\nu \sim \frac{\Omega_{d-3}}{d-3}\left ( -g^{\mu\nu} - \E_1^\mu \E_1^\nu \right)\, ,
\end{align}
where $\sim$ means that we are dropping terms which are vanishing when contracted with the $\T^{\mu\nu}_{\A}$ tensor, due to gauge invariance.
Taking into account the previous considerations, we can rewrite $\I_\Phi^{\mu\nu}$ as
\begin{align}
     \I_\Phi^{\mu\nu} \! &=
    \frac{\Omega_{d-3}}{\Omega_{d-2}} 
 \!\int_0^\pi \!\!\!d \Phi(\sin\Phi)^{-2\epsilon} g(\Phi)\!
 \left( \!- g^{\mu\nu} \!\! -2(1-\epsilon) \cos^2\!\Phi\, {\E_1}^\mu{\E_1}^\nu\! -\frac{2-2\epsilon}{1-2\epsilon}\sin^2\!\Phi  \left ( \!-g^{\mu\nu} \!\!- \!\E_1^\mu \E_1^\nu \right)  \!\!\right) \notag \\
 &= \frac{\Omega_{d-3}}{\Omega_{d-2}} 
 \int_0^\pi d \Phi(\sin\Phi)^{-2\epsilon}\, g(\Phi)
\left( \cos^2\!\Phi - \frac{\sin^2\!\Phi}{1-2\epsilon} \right) \biggl( (-g^{\mu\nu} +\Amunu)\epsilon  -\Amunu \biggr)
\end{align}
where we introduced the asymmetry tensor $\Amunu= 2\E_1^\mu \E_1^\nu +g^{\mu\nu}$.
It is worth noticing that, if $g(\Phi)$ is constant, the integral $\I_\Phi^{\mu\nu}$ is identically zero to all orders in $\epsilon$.

In the full computation of the initial- and final-state collinear contributions, $\I_\Phi^{\mu\nu}$ multiplies a single $1/\epsilon$ pole and, thus, we need the result up to $\mathcal{O}(\epsilon^2)$, namely
\begin{align}
	\I_\Phi^{\mu\nu} = \begin{cases}
	 -\int_0^\pi \frac{d \phi}{\pi} \left(\frac{1+\cos\phi}{\sin\phi}\right)^{2\epsilon} \left( \cos^2\phi - \frac{\sin^2\phi}{1-2\epsilon} \right)\Amunu +\mathcal{O}(\epsilon^2) 
	 = \epsilon \Amunu +\mathcal{O}(\epsilon^2) 
	 & \text{for ISC} \\
	 - \int_0^\pi \frac{d \varphi}{\pi} \left( \cos^2\varphi - \frac{\sin^2\varphi}{1-2\epsilon} \right)\Amunu +\mathcal{O}(\epsilon^2)    
	 = \epsilon \Amunu +\mathcal{O}(\epsilon^2) 
	 &\text{for FSC}
	\end{cases}\, .
\end{align}
In conclusion, it turns out that the spin-dependent part of the $\I_{*}$ integrals is proportional to the contraction $\Amunu \T_{\A,\mu\nu}$. This quantity can be directly evaluated in $d=4$ dimensions as
\begin{align}
  \label{eq:AmunuTmunu}
	\Amunu \T_{\A, \mu\nu} = |\mathcal{M}_{\A}^{(0),\E_{1}}|^2
	- |\mathcal{M}_{\A}^{(0),\E_2}|^2
\end{align}
where we defined the polarised Born matrix elements as 
$ |\mathcal{M}_{\A}^{(0),\E_i}|^2=\T_{\A, \mu\nu}\E_i^\mu \E_i^\nu$, $i=1,2$. 
Having these results in mind, we can derive the formul\ae\ in Eqs. (\ref{eq:I_ISC_DET_result}) and (\ref{eq:I_FSC_DET_result}).

\section{Splitting kernels}
\label{sec:APkernels}
In the initial-state collinear limit, we use the regularised and spin-polarised splitting kernels
\begin{align}
	P_{qq}^{ss'}(z,\hat{k}_t;\epsilon) &= P_{qq}(z;\epsilon)\,\delta^{ss'} \notag\\
	P_{qg}^{ss'}(z,\hat{k}_t;\epsilon)  &= P_{qg}(z;\epsilon) \,\delta^{ss'}  \notag\\
	P_{gg}^{\mu\nu}(z,\hat{k}_t;\epsilon)  &= P_{gg}(z;\epsilon)(-g^{\mu\nu}) - G_{gg}(z) \biggl( -g^{\mu\nu} - 2(1-\epsilon)\hat{k}_t^{\mu}\hat{k}_t^{\nu}  \biggr)  \notag\\
	P_{gq}^{\mu\nu}(z,\hat{k}_t;\epsilon)   &=  P_{gq}(z;\epsilon)(-g^{\mu\nu}) - G_{gq}(z) \biggl( -g^{\mu\nu} - 2(1-\epsilon)\hat{k}_t^{\mu}\hat{k}_t^{\nu}   \biggr) 
\end{align}
where the spin-averaged splitting kernels are
\begin{align}
	P_{qq}(z;\epsilon) &= \frac{C_F}{2}\left( \frac{z^2+1}{(1-z)_+} -\epsilon(1-z) \right)+\frac{3}{4}C_F\delta(1-z) \notag \\
	P_{qg}(z;\epsilon) &= \frac{T_R}{2}\left( 1-\frac{2z(1-z)}{1-\epsilon} \right) \notag \\
	P_{gg}(z;\epsilon) &= C_A\left( \frac{z}{(1-z)_+} + \frac{1-z}{z} + z(1-z)  \right) +\beta_0\delta(1-z) \notag \\
	P_{gq}(z;\epsilon) &= \frac{C_F}{2}\left( \frac{(1-z)^2+1}{z} -\epsilon z \right) \, .
\end{align}
One can obtain the unregularised splitting kernels $\hat{P}$ by simply replacing $(1-z)_+\to (1-z)$ and dropping terms proportional to $\delta(1-z)$.
The functions $G_{ab}(z)$ are \cite{Catani:2010pd}
\begin{align}
	G_{gg}(z) = C_A \frac{1-z}{z}  &&
	G_{gq}(z) =  C_F \frac{1-z}{z}  \,
\end{align}
and the functions $C_{ab}(z)$ read \cite{Collins:1984kg}
\begin{align}
C_{gq}(z) &= \frac{C_F}{2}z  & C_{qq}(z) &= \frac{C_F}{2}(1-z) & C_{qg}(z) = T_R z(1-z)\, .
\end{align}
In the final-state collinear limit, we use the unregularised and spin-polarised splitting kernels
\begin{align}
	\hat{P}_{q \to qg}^{ss'}(\xi,\hat{k}_\bot;\epsilon) &= \hat{P}_{q\to qg}(\xi;\epsilon)\,\delta^{ss'} = \hat{P}_{q\to gq}^{ss'}(1-\xi,k_\bot;\epsilon)  \notag\\
	\hat{P}_{g \to gg}^{\mu\nu}(\xi,\hat{k}_{\bot};\epsilon)  &= \hat{P}_{g \to gg}(\xi;\epsilon)(-g^{\mu\nu}) - G_{g \to gg}(\xi,\epsilon) \biggl( -g^{\mu\nu} - 2(1-\epsilon)\hat{k}_\bot^{\mu}\hat{k}_\bot^{\nu}  \biggr)  \notag\\
	\hat{P}_{g \to q\bar{q}}^{\mu\nu'}(\xi,\hat{k}_{\bot};\epsilon)   &=  \hat{P}_{g \to q\bar{q}}(\xi;\epsilon)(-g^{\mu\nu}) - G_{g \to q\bar{q}}(\xi,\epsilon) \biggl( -g^{\mu\nu} - 2(1-\epsilon)\hat{k}_\bot^{\mu}\hat{k}_\bot^{\nu}  \biggr) 
\end{align}
where the spin-averaged splitting kernels and the functions $G_{g\to gg}(\xi,\epsilon)$ and $G_{g\to q{\bar q}}(\xi,\epsilon)$ read
\begin{align}
	\hat{P}_{q \to qg}(\xi;\epsilon) &= \frac{C_F}{2}\left( \frac{\xi^2+1}{1-\xi} -\epsilon(1-\xi) \right) = \hat{P}_{q \to gq}(1-\xi;\epsilon)  \notag \\
	\hat{P}_{g \to gg}(\xi;\epsilon) &= \frac{C_A}{2}\left( \frac{\xi}{1-\xi} + \frac{1-\xi}{\xi} + \xi(1-\xi)  \right) \notag \\
	\hat{P}_{g \to q\bar{q}}(\xi;\epsilon) &= \frac{T_R}{2}\left( 1-\frac{2\xi(1-\xi)}{1-\epsilon}\right) \notag \\
	G_{g \to gg}(\xi,\epsilon) &= \frac{C_A}{2} \xi(1-\xi) \notag \\
	G_{g \to q\bar{q}}(\xi,\epsilon) &=  -\frac{T_R}{1-\epsilon} \xi(1-\xi)  \,.
\end{align}
Note that we already include a symmetry factor $1/2$ for the identical gluons in the final-state $g\to gg$ splitting kernels and $G$ functions, i.e., there are no symmetry factors in the phase space. 

\section{Soft integrals for $\ktness$}
\label{sec:bspace_integral}
In this Appendix we discuss the integration, over the radiation phase space, of the soft singular term $\mathbf{J}^2_{\rm sing} $ defined in Eq.~\eqref{eq:J_sing}. 
We start by noticing that the result is the sum of integrals of the type
\begin{equation} \label{eq:J_integral_below_cut}
	2 \mu_R^{2 \epsilon} \frac{e^{\gamma_E\epsilon}}{\Gamma(1-\epsilon)} \int \frac{d^d k}{\Omega_{d-2}} \,\delta_{+}(k^2) \Theta\left(\rcut-\frac{k_T}{Q}\right) J(k, p; v_1,v_2),
\end{equation}
where the function $J(k, p; v_1,v_2) =\frac{1}{p \cdot k}\left(\frac{p \cdot v_1}{v_1 \cdot k}-\frac{p \cdot v_2}{v_2 \cdot k}\right)$ is integrated over the radiation phase space $d^d k$ and depends on two massive four-vectors $v_i$ $(i=1,2)$ and a massless four-vector $p$.
Indeed, we can consider Eq.~\eqref{eq:J_sing} and easily identify three families of integrals of the type in Eq.~\eqref{eq:J_integral_below_cut} according to the choice of $p$, $v_1$ and $v_2$.\
The first contribution is 
\begin{equation}
	\omega^1_2 - \omega^1_i=\frac{1}{p_1 \cdot k}\left(\frac{p_1 \cdot\left(p_1+p_2\right)}{k \cdot\left(p_1+p_2\right)}-\frac{p_1 \cdot\left(p_1+p_i\right)}{k \cdot\left(p_1+p_i\right)}\right) = J(k, p_1; p_1+p_2, p_1+p_i)  \,,
\end{equation}
where $p_1$ and $p_2$ are initial-state momenta and $p_i$ is the momentum of a final-state parton.
The second contribution is of type 
\begin{equation}
	\omega_{C_i,S} - \omega^i_1 =\frac{1}{p_i \cdot k}\left(\frac{p_i \cdot\left(p_1+p_2\right)}{k \cdot\left(p_1+p_2\right)}-\frac{p_i \cdot\left(p_i+p_1 \right)}{k \cdot\left(p_i+p_1\right)}\right) = J(k, p_i; p_1+p_2, p_i+p_1) \,,
\end{equation}
while the last contribution is of type 
\begin{equation}
	\omega_{C_i,S} - \omega^i_j =\frac{1}{p_i \cdot k}\left(\frac{p_i \cdot\left(p_1+p_2\right)}{k \cdot\left(p_1+p_2\right)}-\frac{p_i \cdot\left(p_i+p_j \right)}{k \cdot\left(p_i+p_j\right)}\right)= J(k, p_i; p_1+p_2, p_i+p_j).
\end{equation}
The $b$-space transformation for the function $J(k, p; v_1,v_2)$ is known from $q_T$-resummation of heavy-quark pair production and it is given by \cite{Catani:2023tby}
\begin{equation}
\label{eq:b_space}
\begin{aligned}
	\langle I ({\vec b})\rangle &=\left\langle\int d^d k \delta_{+}\left(k^2\right) e^{i \vec{k}_T \cdot \vec{b}}\, J(k, p; v_1,v_2) \right\rangle \\
	&=\frac{1}{4} \Omega_{d-2} \Gamma^2(1-\epsilon)\left(\frac{b^2}{4}\right)^\epsilon \, \!\!\left[-\frac{2}{\epsilon} \ln\!\left(\frac{p \cdot v_1 \sqrt{v_2^2}}{p \cdot v_2 \sqrt{v_1^2}}\right)\!+\!\mathrm{Li}_2\!\left(\!-\frac{v_{1, T}^2}{v_1^2}\right)\!-\!\mathrm{Li}_2\!\left(\!-\frac{v_{2, T}^2}{v_2^2}\right)\!+\!\mathrm{O}(\epsilon)\right] ,
\end{aligned}
\end{equation}
where $b^2 \equiv (\vec{b})^2$, $v_{i,T}^2 \equiv (\vec{v}_{i,T})^2$ and $\langle \cdots \rangle$ stands for the $(d-2)$-dimensional azimuthal average over the direction of the impact parameter $\vec{b}$. 

The integral in Eq. \eqref{eq:J_integral_below_cut} can be related to Eq. \eqref{eq:b_space} via
\begin{align}
	\label{eq:b-space-to-cum-space}
	2 \mu_R^{2 \epsilon} \frac{e^{\gamma_E\epsilon}}{\Gamma(1-\epsilon)} \int \frac{d^d k}{\Omega_{d-2}} \,\delta_{+}(k^2) &\Theta\left(\rcut-\frac{k_T}{Q}\right) J(k, p; v_1,v_2) =\notag\\
	&=\left(\frac{\mu_R^2 }{Q^2}\right)^\epsilon \rcut^{-2\epsilon}  \frac{e^{\gamma_E \epsilon}}{\Gamma(1-\epsilon)} \frac{2 \langle I({\vec b})\rangle}{\Omega_{d-2} \Gamma^2(1-\epsilon)\left(\frac{b^2}{4}\right)^\epsilon}\,,
\end{align}
where $Q = \sqrt{(p_1 + p_2)^2}$ is the invariant mass of the system. The previous relation can be derived by writing the $\Theta$-function in Fourier space as 
\begin{equation}
	\Theta\left(\rcut-\frac{k_T}{Q}\right) = \int_0^{\rcut Q} d^{d-2}k'_T\, \delta^{(d-2)}(\vec{k}_T-\vec{k'}_T) = \int_0^{\rcut Q} d^{d-2}k'_T\, \int \frac{d^{d-2}b}{(2\pi)^{d-2}} e^{i (\vec{k}_T - \vec{k'}_T)\cdot \vec{b}}
\end{equation}
and then performing the integral over the $(d-2)$-dimensional azimuthal space around the direction of the impact parameter.
The final result for the integral in Eq. (\ref{eq:J_integral_below_cut}) depends on the transverse momenta $v_{i, T}$ with respect to the massless four-vector $p$.
The transverse component $v_T$ of a generic four-vector $v$ can be defined uniquely by introducing a generic reference vector $N$ and by considering the following decomposition:
\begin{equation}
	v_T=v-\frac{v\cdot \left(N-p\frac{N^2}{N\cdot p}\right)}{N\cdot p}p-\frac{v\cdot p}{N \cdot p}N\,, 
\end{equation}
which satisfies the conditions $p\cdot v_T= N\cdot v_T=0$.
In this paper we implicitly choose $N=p_1+p_2$ such that $\vec{v}_T \equiv \vec{v}_t$ coincides with the transverse momentum with respect to the beam direction if $p$ is an initial-state parton, while $\vec{v}_T \equiv \vec{v}_{\perp}$ lives in the plane orthogonal to $\vec{p}$ in the Born CM frame (note that in this frame $\vec p_1+\vec p_2=\vec 0$).

Finally, we obtain
\begin{align}
	2 \mu_R^{2 \epsilon} \frac{e^{\gamma_E\epsilon}}{\Gamma(1-\epsilon)} &\int \frac{d^d k}{\Omega_{d-2}} \,\delta_{+}(k^2) \Theta\left(\rcut-\frac{k_t}{Q}\right) J(k, p_1; p_1+p_2, p_1+p_i)  \notag\\
&=\frac{1}{2} \left(\frac{\mu_R^2 }{Q^2}\right)^{\epsilon} \rcut^{-2\epsilon} \,\frac{ e^{\gamma_E \epsilon}}{\Gamma(1-\epsilon)} \left\{ -\frac{1}{\epsilon} \ln \left(\frac{p_1 \cdot p_2}{p_1 \cdot p_i}\right) +\mathrm{ Li}_2\left(-\frac{p_{i, t}^2}{2 p_1 \cdot p_i}\right) \right\} +\mathrm{O}(\epsilon) ,
\end{align}
\begin{align}
	&2 \mu_R^{2 \epsilon} \frac{e^{\gamma_E\epsilon}}{\Gamma(1-\epsilon)}  \int \frac{d^d k}{\Omega_{d-2}} \,\delta_{+}(k^2) \Theta\left(\rcut-\frac{k_\perp}{Q}\right)J(k, p_i; p_1+p_2, p_i+p_1)  \notag\\
	&=\frac{1}{2} \left(\frac{\mu_R^2 }{Q^2}\right)^{\epsilon}  \rcut^{-2\epsilon}  \,\frac{e^{\gamma_E \epsilon}}{\Gamma(1-\epsilon)} \left\{-\frac{2}{\epsilon} \ln \left(\frac{p_i \cdot (p_1 +p_2) \sqrt{2 p_i \cdot p_1}}{p_i \cdot p_1 \sqrt{2 p_1 \cdot p_2}}\right) +\operatorname{Li}_2\left(-\frac{p_{1, \perp}^2}{2 p_1 \cdot p_i}\right)\right\} +\mathrm{O}(\epsilon) \notag\\
	&= \frac{1}{2} \left(\frac{ \mu_R^2 }{Q^2}\right)^{\epsilon} \rcut^{-2\epsilon}  \,\frac{ e^{\gamma_E \epsilon}}{\Gamma(1-\epsilon)} \left\{-\frac{1}{\epsilon} \ln \left( \frac{2 E^2_{J_i}}{ p_i \cdot p_1}\right) +\operatorname{Li}_2\left(-\frac{p_{1, \perp}^2}{2 p_1 \cdot p_i}\right)\right\} +\mathrm{O}(\epsilon),
\end{align}
\begin{align}
	&2 \mu_R^{2 \epsilon} \frac{e^{\gamma_E\epsilon}}{\Gamma(1-\epsilon)}   \int \frac{d^d k}{\Omega_{d-2}} \,\delta_{+}(k^2) \Theta\left(\rcut-\frac{k_\perp}{Q}\right)J(k, p_i; p_1+p_2, p_i+p_j)  \notag \\
	&=\frac{1}{2} \left(\frac{\mu_R^2 }{Q^2}\right)^{\epsilon} \rcut^{-2\epsilon}\frac{e^{\gamma_E \epsilon}}{\Gamma(1-\epsilon)} \left\{-\frac{2}{\epsilon} \ln \left(\frac{p_i \cdot (p_1 +p_2) \sqrt{2 p_i \cdot p_j}}{p_i \cdot p_j \sqrt{2 p_1 \cdot p_2}}\right) +\operatorname{Li}_2\left(-\frac{p_{j, \perp}^2}{2 p_j \cdot p_i}\right)\right\} +\mathrm{O}(\epsilon) \notag\\
	&= \frac{1}{2} \left(\frac{\mu_R^2 }{Q^2}\right)^{\epsilon} \rcut^{-2\epsilon}\frac{e^{\gamma_E \epsilon}}{\Gamma(1-\epsilon)} \left\{-\frac{1}{\epsilon} \ln \left( \frac{2 E^2_{J_i}}{ p_i \cdot p_j}\right) +\operatorname{Li}_2\left(-\frac{p_{j, \perp}^2}{2 p_j \cdot p_i}\right)\right\} +\mathrm{O}(\epsilon)\, ,
\end{align}
for the three families of integrals introduced above.

\section{SCET-like definition of the NLO jet function}
\label{app:Scet-like-jet-function}

In Sect.~(\ref{sec:FSC}) we obtained the contribution from the final-state collinear radiation by starting from an exact parametrization of the phase space. Then, we expanded the matrix element and parts of the phase space in the limit where the angle between the two collinear partons becomes small. The advantage of this method is that it yields jet and soft functions with a very clear physical origin. The jet function always contains the collinear singularity as well as the soft-collinear contribution; its definition does not depend on the scaling of the slicing variable in the respective region and, in particular, it is agnostic to whether the slicing variable belongs to the class of SCET$_{\rm I}$ or SCET$_{\rm II}$ problems.
The (subtracted) soft function is then only related to soft wide-angle radiation.

On the other hand the method outlined in Sect.~(\ref{sec:FSC}) has the disadvantage that it does not achieve a full separation of scales between the collinear and the soft regions, even in cases where such a separation is possible. To see this, consider the result of Eq.~\eqref{eq:I_FSC_DET_result} for the final-state collinear region in the case of $\DeltaET$-slicing. The result depends non-trivially on the jet energy  $E_J$ and on the jet transverse momentum $p_{J,t}$. However, when the final-state collinear region is combined with the soft one in Sect.~(\ref{sec:DeltaETSigmas}), the dependence on $E_J$ drops out, suggesting that it is unphysical and frame dependent. To make the separation of scales more apparent, we can follow a strategy similar to the one exploited in SCET or, in other words, we can exploit the method of regions to give a definition of the jet and soft functions.

In order to identify the different regions, we need to introduce collinear and anti-collinear reference vectors for each (Born level) jet. 
We find it useful to set up the reference vectors in such a way that the transverse components of the momenta are purely spatial in the partonic CM frame. To this purpose, we introduce a massive reference vector $N=p_1+p_2$, where $p_1$ and $p_2$ are the momenta of the initial-state partons. Having this in mind, for a given (massless) Born level jet with four-momentum $p$, we can define the collinear and anti-collinear reference vectors
\begin{align}
	n&=\frac{\sqrt{N^2}}{p\cdot N}p \,,& \bar{n}=&\frac{2}{\sqrt{N^2}}N-n
\end{align}
satisfying $n^2=\Bar{n}^2=0$ and $n\cdot \Bar{n}=2$. Then, we can decompose every four-vector $k$ as 
\begin{align}
	k=k^- n + k^+ \Bar{n} +k_\perp\, ,
\end{align}
where $k^-=\frac{k\cdot  \Bar{n}}{2}$ and $k^+=\frac{k\cdot n}{2}$, and $k_\perp$ is the purely spatial transverse component with $k_\perp \cdot n = k_\perp \cdot \bar{n} = 0$. We also introduce the collinear momentum fraction $z=\frac{k \cdot \Bar{n} }{p\cdot \Bar{n}}=\frac{k_-}{p_-}$.

We consider the splitting $a_i(\tilde{p}_i) \rightarrow a(k) + b(p_i)$. For a generic resolution variable $r$, we can define the differential NLO jet function as
\begin{align}
	\label{eq:SCET-like-Jetfunction}
	{\tilde \jmath}_{a_i \to ab}^{\,ss'}(r) &=
        \frac{\mu_R^{2\epsilon}}{\Omega_{2-2\epsilon}}\frac{e^{\epsilon\gamma_E}}{\Gamma(1-\epsilon)}
        \frac{\as}{\pi} \int \frac{dz_{k}}{z_k}\int \frac{dz_{p_{i}}}{z_{p_{i}}}\int d^{d-2} k_{\perp}\int d^{d-2} p_{\perp,i}\frac{\hat{P}_{a_i\to ab}^{ss'}(\{k\})}{k \cdot p_i}  \notag \\
	&\hspace{1.5cm}\times \delta(z_k+z_{p_{i}}-1)\delta^{d-2}(k_{\perp}+p_{\perp,i})\delta(r -r^{C_i}(\{p_3,\ldots,\tilde{p}_{i},\ldots,p_{n+2}\}))\notag \\
        &= \mu_R^{2\epsilon} \frac{e^{\epsilon\gamma_E}}{\Gamma(1-\epsilon)}
        \frac{\as}{\pi} \int \frac{d\Omega_{2-2\epsilon}}{\Omega_{2-2\epsilon}}\int_0^1 \! dz\! \int\! \frac{d k_{\perp}^2}{(k_{\perp}^2)^{1+\epsilon}}\hat{P}_{a_i\to ab}^{ss'}(z,\hat{k}_{\perp};\epsilon) \delta(r  -r^{C_i}(z,k_{\perp})) \notag \\
        & =\left(\frac{\mu_R^2}{Q^2}\right)^\epsilon \frac{e^{\epsilon\gamma_E}}{\Gamma(1-\epsilon)}
        \frac{\as}{\pi} \int \frac{d\Omega_{2-2\epsilon}}{\Omega_{2-2\epsilon}}\int_0^1 dz z^{-\epsilon}(1-z)^{-\epsilon} \notag \\
        &\hspace{4cm} \times \int d x x^{-1-\epsilon}\hat{P}_{a_i\to ab}^{ss'}(z,\hat{k}_{\perp};\epsilon) \delta(r  -r^{C_i}(z,x))\,,
\end{align}
where $\hat{P}_{a_i\to ab}^{ss'}$ is the unregularised and spin-polarised splitting kernel (Appendix \ref{sec:APkernels}), $r^{C_i}$ is the approximation of  the slicing variable in the collinear region defined above and $Q^2$ is the squared CM energy. In the last step we have performed the change of variables to the dimensionless invariant mass $ x = 2 k\cdot p_{i}/Q^{2}= k_{\perp}^{2}/(Q^{2}z(1-z))$ at fixed $z$.
The cumulant version of the NLO jet function is obtained by integrating the differential jet function up to $\rcut$ and summing over all possible splittings of parton $a_i$
\begin{align}
	\mathcal{\widetilde J}_{a_i}^{ss'}(\rcut) = \sum_{(*)} \,\int_0^{\rcut} dr \, {\tilde \jmath}_{a_i \to (*)}^{\,ss'}(r) \,,
\end{align}
where $(*)$ labels an arbitrary splitting.\\
Thus, the final-state collinear contribution associated with the limit $k \cdot p_i \rightarrow 0$ is given by
\begin{align}
	\widetilde{\FSC_i} &= \sum_{\A}\int_0^1 dx_1 f_{a_1}(x_1,\mu_F)\int_0^1 dx_2 f_{a_2}(x_2,\mu_F) \int\! d\Pi^{d}_n(q;p_F,\{p_j\}_{j=3}^{n+2}) \frac{\T_{a_1a_2;\dots a_i\dots}^{ss'}}{2Q^2} \,\mathcal{\widetilde J}_{a_i}^{ss'}(\rcut),
\end{align}
where $q = p_1 + p_2$ and we summed over the possible Born configurations $\A = \{a_1, a_2, \{a_i\}\}$.\\
By inspecting Eq.~\eqref{eq:SCET-like-Jetfunction}, the new definition of the NLO jet function differs from the one in Eq.~(\ref{eq:cum_jet_function}) by the absence of the phase space factor beyond the strictly collinear limit given in Eq.~\eqref{eq:ps-factor-approx}. 
Even though it seems that the soft endpoints $z\to 0$ and $z\to 1$ are regulated by the factors $z^{-\epsilon}$ and $(1-z)^{-\epsilon}$ respectively, the integral over the variable $x$ can generate additional $z$-dependent terms that spoil this regularisation.
Therefore, the SCET-like definition leads to the possible appearance of rapidity divergences as expected when applying the method of regions. One way to treat these rapidity divergences is to consistently introduce rapidity regulators \cite{Ji:2004wu,Chiu:2009yx,Becher:2011dz,Chiu:2012ir,Echevarria:2015byo,Li:2016axz,Chay:2020jzn} in the jet, beam and soft functions associated with each singular region. To make contact with the definition given in Sect.~\ref{sec:FSC}, we consider here a different method inspired by Ref.~\cite{Catani:2022sgr}.

We modify the jet function of Eq.~(\ref{eq:SCET-like-Jetfunction}) by applying the replacement
\begin{equation}
  z=\frac{k\cdot \bar{n}}{\tilde{p}_{i}\cdot \bar{n}}= \frac{k\cdot\Bar{n}}{(k+p_i)\cdot \Bar{n}} \rightarrow  z_{N}=\frac{k\cdot N}{(k+p_i)\cdot N}
\end{equation}
in the argument of the splitting kernel $\hat{P}_{a_i\to ab}^{ss'}(z,\hat{k}_{\perp};\epsilon)$. We refer to this procedure as the ``$z_N$-prescription". We observe that the fraction $z_{N}$ with the choice $N=p_{1}+p_{2}$ coincides with the energy fraction $\xi$ defined in Sect.~\ref{sec:FSC}. At fixed $x$, we can express $z$ in terms of $z_{N}$ as
\begin{equation}\label{eq:zzNrel}
  z \simeq \frac{z_{N}}{1-Q^{2}x/4 (\tilde{p}^{-}_{i})^{2}} \left(1 - \frac{Q^{2}}{4(\tilde{p}^{-}_{i})^{2}}\frac{x(1-z_{N})}{z_{N}}\right)\,.
\end{equation}
Performing a change of variables from $z$ to $z_{N}$ at fixed $x$, the NLO differential jet function with the $z_{N}$-prescription becomes
\begin{align}
  {\tilde \jmath}_{a_i \to ab}^{\,ss'}(r) =&
  \left(\frac{\mu_R^2}{Q^2}\right)^\epsilon\! \frac{e^{\epsilon\gamma_E}}{\Gamma(1-\epsilon)}
  \frac{\as}{\pi} \int \frac{d\Omega_{2-2\epsilon}}{\Omega_{2-2\epsilon}}\int_0^1  dz z^{-\epsilon}(1-z)^{-\epsilon}\notag\\
  & \hspace{0.5cm}\times\int d x x^{-1-\epsilon}\hat{P}_{a_i\to ab}^{ss'}(z_{N},\hat{k}_{\perp};\epsilon) \delta(r  -r^{C_i}(z,x))  \notag \\
   = &  \left(\frac{\mu_R^2}{Q^2}\right)^\epsilon\! \frac{e^{\epsilon\gamma_E}}{\Gamma(1-\epsilon)}\frac{\as}{\pi} \int \frac{d\Omega_{2-2\epsilon}}{\Omega_{2-2\epsilon}} \int_0^1 dz_{N} z_{N}^{-\epsilon}(1-z_{N})^{-\epsilon}  \notag \\ 
   & \hspace{0.5cm} \times \!  \int \! d x x^{-1-\epsilon}\! \left(1-\frac{Q^{2}}{4(\tilde{p}^{-}_{i})^{2}}\frac{x}{z_{N}(1-z_{N})}\right)^{-\epsilon}\!\!\hat{P}_{a_i\to ab}^{ss'}(z_{N},\hat{k}_{\perp};\epsilon) \delta(r  -r_{N}^{C_i}(z_{N},x)),
\end{align}
where we have introduced the notation $r_{N}^{C_i}(z_{N},x):=r^{C_i}(z(z_{N},x),x)$. In the final expression of the above equation, we are allowed to identify $\tilde{p}^{-}_{i}\approx \tilde{p}^{0}_{i}$, where $\tilde{p}^{0}_{i}$ is the energy component in the partonic CM frame. The effect of the $z_{N}$-prescription is that of reinstating power corrections beyond the strictly collinear limit. More precisely, it allows us to exactly recover the additional phase space factor in Eq.~(\ref{eq:xidefinition}) starting from a careful expansion in the collinear and soft limits. Therefore, the SCET-like jet function regularised with the $z_{N}$-prescription is equivalent to the one obtained in Sect.~\ref{sec:FSC} up to the precise expression used for the approximation of the observable in the collinear limit, $r_{N}^{C_i}$. To be concrete, we consider the case of $\ktness$ and define
$\left(k_{T}^{{\rm ness},C_{i}}(z,x)\right)^{2} = Q^{2}z x / ((1-z)D^{2})$, limiting ourselves, for simplicity, to the case $z<1/2$ (see Eq.~(\ref{eq:ktenssFSCapprox})). Then, applying the change of variables in Eq.~\eqref{eq:zzNrel}, we get
\begin{align}
  \left(k_{T,N}^{{\rm ness},C_{i}}(z_{N},x)\right)^{2} &= \left(k_{T}^{{\rm ness},C_{i}}(z_{N},x)\right)^{2}\frac{1-\frac{Q^{2}}{4(\tilde{p}^{-}_{i})^{2}}\frac{x(1-z_{N})}{z_{N}}}{1-\frac{Q^{2}}{4(\tilde{p}^{-}_{i})^{2}}\frac{x z_{N}}{1-z_{N}}} \notag \\ &\approx \left(k_{T}^{{\rm ness},C_{i}}(z_{N},x)\right)^{2}\left( 1-\frac{Q^{2}}{4(\tilde{p}^{-}_{i})^{2}}\frac{x(1-2z_{N})}{z_{N}(1-z_N)} \right)\notag\\
  &\approx\left(k_{T}^{{\rm ness},C_{i}}(z_{N},x)\right)^{2}\left( 1-\frac{Q^{2}}{4(\tilde{p}^{-}_{i})^{2}}\frac{x}{z_{N}(1-z_N)} \right).
\end{align}
On the other hand, in Sect.~\ref{sec:FSC_ktness}, we made the choice $\left(k_{T,N}^{{\rm ness},C_{i}}(z_{N},x)\right)^{2} \!\!=\!  \left(k_{T}^{{\rm ness},C_{i}}(z_{N},x)\right)^{2}$, see Eq.~\eqref{eq:ktenssFSCapprox}. As we will show explicitly in the following, this does not change the coefficients of the poles in $d$ dimensions nor those of the logarithms of the observable. Only the constant term can be affected. The difference is compensated by the corresponding change in the soft function. Indeed, the modification of the observable in the collinear limit affects also the related  subtraction contribution in the definition of the subtracted current in Eq.~\eqref{eq:J2sub}.
More precisely, the soft limit of the SCET-jet function regularised with the $z_{N}$-prescription reads
\begin{equation}
		\lim_{k^{0} \to 0} {\tilde \jmath}_{a_i \to ab}^{\,ss'}(r) = \mu_R^{2\epsilon}\frac{e^{\epsilon\gamma_E}}{\Gamma(1-\epsilon)} \frac{\as}{\pi} C_{a_{i}}\int \frac{d^dk}{\Omega_{d-2}}\delta_+(k^2)\frac{2\tilde{p}_i\cdot N}{(k\cdot \tilde{p}_i)( k\cdot N)}\delta(r-r^{C_{i}S}(k))\,,
\end{equation}
where $C_{a_i}$ is the colour Casimir of the parton $a_{i}$ and $r^{C_{i}S}(k)$ is obtained by first taking the collinear limit and then the limit where $k$ becomes soft, i.e. taking the leading contribution of $r^{C_i}(z,x)$ for $z \to 0$.
In the case of $\ktness$, this translates into
\begin{equation}
  \left(\frac{k_{T}^{{\rm ness},C_{i}S}}{Q^{2}}\right)^{2}  = \frac{z x}{D^2} \approx z_N\left(1-\frac{Q^2}{4(\tilde{p}^-_i)^2}\frac{x}{z_N}\right)\frac{x}{D^2}\approx
  \frac{k^{0}}{\tilde{p}^{0}_{i}}\frac{2 k \cdot p_{i}}{Q^{2}D^{2}}\left(1-\frac{k \cdot p_{i}}{2\tilde{p}^{0}_{i}k^{0}}\right)\,,
\end{equation}
to be compared with Eq.~\eqref{eq:ktness2FSCS}.

To conclude, in this Appendix we have seen that a SCET-like definition of the jet function leads to a clean separation of the different scales.
On the other hand, if rapidity divergences are present, the jet function becomes dependent on the scheme chosen to regularise them.
Moreover, the regularisation procedure will spoil such a simple scale separation.
The ensuing scaling violations, however, are typically easy to extract, at least at NLO (see Eq.~(\ref{eq:zN_to_Plus_prescr}) below).
In the following, we will compute the jet function in the SCET-like formulation for the two resolution variables considered in this paper, namely $\DeltaET$ and $\ktness$. 
We will also directly compare the jet functions obtained applying the $z_{N}$-prescription with those computed in Sect.~\ref{sec:FSC_DeltaET} and Sect.~\ref{sec:FSC_ktness}, respectively.

\subsection{$\DeltaET$ jet function}

In the $\DeltaET$ case, no rapidity divergences arise, and there is in principle no need to use any regularisation procedure.
The SCET-like $\DeltaET$ jet function, at the differential level, is given by
\begin{align}
	{\tilde \jmath}_{a_i \to ab}^{\,ss'}(\DeltaET/Q) &=  \left(\frac{\mu_R^2}{Q^{2}}\right)^\epsilon\frac{e^{\epsilon\gamma_E} }{\Gamma(1-\epsilon)} \frac{\as}{\pi}\int_0^1 dz z^{-\epsilon}(1-z)^{-\epsilon} \int dx x^{-1-\epsilon} \notag \\ & \hspace{4cm} \times \int \frac{d\Omega_{2-2\epsilon}}{\Omega_{2-2\epsilon}} \hat{P}_{a_i\to ab}^{ss'}(z,k_{\perp};\epsilon) \,\delta\left(\frac{\DeltaET}{Q} - \frac{Q x \sin^2\phi}{2\tilde{p}_{i,t}}\right) \notag \\
	&\!\!\!\!\!\!\!\!= \frac{\left( \frac{\mu_R^2}{2 \DeltaET \tilde{p}_{i,t}}\right)^{\epsilon}}{\DeltaET/Q} \,\frac{e^{\epsilon\gamma_E}}{\Gamma(1-\epsilon)} \frac{\as}{\pi}\frac{\Omega_{1-2\epsilon}}{\Omega_{2-2\epsilon}}
	\int_0^1 \!\! dz z^{-\epsilon}(1-z)^{-\epsilon} \!\!\int_0^{\pi}\!\! d\phi \hat{P}_{a_i\to ab}^{ss'}(z, \hat{k}_{\perp};\epsilon) \,,
	\label{eq:diff_jetfunction_deltaET}
\end{align}
where $\tilde{p}_{i,t}$ is the transverse momentum with respect to the beam direction.
In the case of a quark-initiated splitting ($a_i=q$), the corresponding kernel does not contain any dependence on the azimuthal angle $\phi$ and the quark jet function is 
\begin{align}
	{\tilde \jmath}_q^{\,ss'}(\DeltaET/Q) &= {\tilde \jmath}_{q\to qg}^{\,ss'}(\DeltaET/Q) \notag \\ &= \delta_{ss'}\,\frac{\left( \frac{\mu_R^2}{8 \DeltaET \tilde{p}_{i,t}}\right)^{\epsilon}}{\DeltaET/Q} \,\frac{e^{\epsilon\gamma_E}}{\Gamma(1-\epsilon)} \frac{\as}{\pi} C_F
	\biggl[ -\frac{1}{\epsilon} -\frac{3}{4} + \epsilon\biggl(-\frac{7}{4} +\frac{\pi^2}{3} \biggr) + \mathcal{O}(\epsilon^2) \biggr]
\end{align}
at the differential level and 
\begin{equation}
	\mathcal{\widetilde J}_q^{ss'}(\rcut) = \delta_{ss'}\left(\frac{\mu_R^2}{Q^2} \right)^\epsilon\frac{e^{\epsilon\gamma_E}}{\Gamma(1-\epsilon)} \frac{\as}{\pi}  \left( \frac{Q}{8 \rcut \tilde{p}_{i,t}}\right)^{\epsilon} C_F
	\biggl[ \frac{1}{\epsilon^2} +\frac{3}{4\epsilon} + \frac{7}{4} -\frac{\pi^2}{3} + \mathcal{O}(\epsilon) \biggr]
\end{equation}
at the cumulant level.
The gluon-initiated ($a_i=g$) splitting kernel contains a residual dependence on the azimuthal angle $\phi$ due to the polarisation of the parent gluon.
By exploiting the technique explained in the Appendix \ref{sec:AzInt}, we can perform the azimuthal integral and obtain
\begin{align}
	{\tilde \jmath}_g^{\,\mu\nu}(\DeltaET/Q) &= {\tilde \jmath}_{g \to gg}^{\,\mu\nu}(\DeltaET/Q) + {\tilde \jmath}_{g \to q \bar{q}}^{\,\mu\nu}(\DeltaET/Q) \notag \\
	&= \frac{\left( \frac{\mu_R^2}{2 \DeltaET \tilde{p}_{i,t}}\right)^{\epsilon}}{\DeltaET/Q} \,\frac{e^{\epsilon\gamma_E}}{\Gamma(1-\epsilon)} \frac{\as}{\pi}\frac{\pi \Omega_{1-2\epsilon}}{\Omega_{2-2\epsilon}}
	\int_0^1 dz z^{-\epsilon}(1-z)^{-\epsilon}  \notag \\
	&\hspace{-2cm}\times \biggl[ -g^{\mu\nu}\biggl(\hat{P}_{g \to gg}(z;\epsilon)+n_f \hat{P}_{g \to q\bar{q}}(z;\epsilon)\biggr) -\epsilon\Amunu \biggl( G_{g \to gg}(z) + n_{f}G_{g \to q\bar{q}}(z) \biggr)+ \mathcal{O}(\epsilon^2)\biggr] \,.
\end{align}
Thus, the gluon jet function is 
\begin{align}
	{\tilde \jmath}_g^{\,\mu\nu}(\DeltaET/Q) = \frac{\left( \frac{\mu_R^2}{8 \DeltaET \tilde{p}_{i,t}}\right)^{\epsilon}}{\DeltaET/Q} \,\frac{e^{\epsilon\gamma_E}}{\Gamma(1-\epsilon)} \frac{\as}{\pi}\biggl\{ &g^{\mu\nu}\biggl[\frac{C_A}{\epsilon} + \beta_0 + C_A \biggl( \frac{67}{36} -\frac{\pi^2}{3}\biggr)\epsilon -T_Rn_f \frac{5}{9}\epsilon \biggr] \notag \\
	&-\epsilon \Amunu\biggl[\frac{C_A}{12}-\frac{T_R n_f}{6} \biggr] + \mathcal{O}(\epsilon^2) \biggr\}		
\end{align}
at the differential level and 
\begin{align}
	\mathcal{\widetilde J}_g^{\mu\nu}(\rcut) &=  \left(\frac{\mu_R^2}{Q^2} \right)^\epsilon\frac{e^{\epsilon\gamma_E}}{\Gamma(1-\epsilon)} \frac{\as}{\pi}   \left( \frac{Q}{8 \rcut \tilde{p}_{i,t}}\right)^{\epsilon} \notag \\ &\times \biggl\{ -g^{\mu\nu}\biggl[\frac{C_A}{\epsilon^2} + \frac{\beta_0}{\epsilon} + C_A \biggl( \frac{67}{36} -\frac{\pi^2}{3}\biggr) -T_Rn_f \frac{5}{9} \biggr]
	+\Amunu\biggl[\frac{C_A}{12}-\frac{T_R n_f}{6} \biggr] + \mathcal{O}(\epsilon) \biggr\}
\end{align}
at the cumulant level.

We observe that the SCET-like jet function does not contain any dependence on the jet energy $E_{J_i}$.
Comparing with the result obtained in Eq. \eqref{eq:cumjetdeltaet} we find
\begin{align}
	 &\mathcal{J}_{a}^{ss'}(\rcut)  - \mathcal{\widetilde J}_{a}^{ss'}(\rcut)  \notag \\ &= d_a^{ss'}\,C_a \left(\frac{\mu_R^2}{Q^2} \right)^\epsilon\frac{e^{\epsilon\gamma_E}}{\Gamma(1-\epsilon)}\frac{\as}{\pi}  \left( \frac{Q}{8 \rcut \tilde{p}_{i,t}}\right)^{\epsilon} \left( \frac{2 E_{J_i}}{Q}\right)^{2 \epsilon}
	\biggr[ -\frac{1}{2\epsilon^2} -\frac{\pi^2}{12} + \mathcal{O}(\epsilon) \biggr]\, .
\end{align}
We see that the two jet functions differ already in the pole structure: this is due to a different partition of contributions of soft and collinear origin between the jet and the soft function.

Even if it is not strictly necessary, it is instructive to consider the result for the SCET-like jet function when applying the $z_N$-prescription. Following the previous discussion, we know that this leads to the jet function in Eq.~(\ref{eq:cum_jet_function}) up to the definition of the observable in the collinear limit. In this case, we have that
\begin{equation}
  \DeltaET^{C}(z,x,\phi) = \frac{Q^{2}x \sin^{2}\phi}{2\tilde{p}_{i,t}} \underset{z \to z(z_{N},x)}{\rightarrow}  = \Delta E_{t,N}^{C}(z_{N},x) = \DeltaET^{C}(z_{N},x)\,,
\end{equation}
since $\DeltaET^{C}(z,x,\phi)$ does not depend on $z$. Therefore, in this case the SCET-like jet function regularised with the $z_N$-prescription coincides exactly with the one defined in the main text. Comparing with the pure SCET-like definition, we have the advantageous feature that the pole structure is observable independent and predictable both in the collinear region and in the pure soft region\footnote{A similar implication was noticed in Ref.~\cite{Bauer:2020npd} when discussing the consequences of introducing a rapidity regulator for SCET$_{\rm I}$ problems.}.
\subsection{$\ktness$ jet function}
Since $\ktness$ behaves as a transverse momentum in every collinear limit, its SCET-like jet function manifests a rapidity divergence, which we regularise with the $z_N$-prescription. At the differential level, we have
\begin{align}
  {\tilde \jmath}_{a_i \to ab}^{\,ss'}(\ktness/Q) &=  \left(\frac{\mu_R^2}{Q^2} \right)^\epsilon\frac{e^{\epsilon\gamma_E}}{\Gamma(1-\epsilon)}\frac{\as}{\pi}\int_0^1 dz z^{-\epsilon}(1-z)^{-\epsilon} \int d x x^{-1-\epsilon} \notag \\
  & \hspace{1.5cm}\times \int\frac{d\Omega_{2-2\epsilon}}{\Omega_{2-2\epsilon}} \hat{P}_{a_i\to ab}^{ss'}(z_N,\hat{k}_{\perp};\epsilon) \,\delta\left(\frac{\ktness}{Q} -\sqrt{\frac{ z(1-z)x}{\max(z,1-z)^{2}D^{2}}}\right) \notag \\
	&=d_{a_i}^{ss'} \frac{\left( \frac{\mu_R^2}{{\ktness}^2D^{2}}\right)^{\epsilon}}{\ktness/Q} \,\frac{e^{\epsilon\gamma_E}}{\Gamma(1-\epsilon)} \frac{\as}{\pi}2
	\int_0^1 dz\max(z,1-z)^{-2\epsilon}\hat{P}_{a_i\to ab}(z_N;\epsilon) \,,
\end{align}
where we used the fact that the slicing variable does not contain any azimuthal dependence which allows us to replace the splitting kernel with its azimuthal average. Starting from Eq.~\eqref{eq:zzNrel}, we express $z_N$ as
\begin{align}
	z_N = \frac{z}{1+\frac{Q^{2}x}{4 (\tilde{p}^{0}_{i})^{2}}}\left(1 + \frac{Q^{2}x}{4 (\tilde{p}^{0}_{i})^{2}}\frac{1-z}{z}\right)
	\approx z\left(1+\frac{{\ktness}^2D^{2}}{4(\tilde{p}^{0}_{i})^{2}z^{2}}\right)\, ,
\end{align}
where the last approximation contains the leading behaviour in the collinear and soft region. At the leading power, we have to perform integrals of the kind
\begin{equation}
  I_{z_{N}} = \int_{0}^{1}d z  \frac{f(z;\epsilon)}{z_{N}}\;,
\end{equation}
where $f(z;\epsilon)$ is finite in the limit $z\to0$. By performing the following manipulation
\begin{eqnarray}
  I_{z_{N}} &=& \int_{0}^{1}d z  \frac{f(z)-f(0)}{z} \frac{1}{1+\frac{{\ktness}^2D^{2}}{4(\tilde{p}^{0}_{i})^{2}z^{2}}} + f(0)\int_{0}^{1}d z \frac{1}{1+\frac{{\ktness}^2D^{2}}{4(\tilde{p}^{0}_{i})^{2}z^{2}}}\notag \\ &=&  \int_{0}^{1}d z  \frac{f(z)-f(0)}{z} -\frac{1}{2}\log\left( \frac{{\ktness}^2D^{2}}{4(\tilde{p}^{0}_{i})^{2}}\right)f(0)+   \mathcal{O}\left(\frac{{\ktness}^2}{(\tilde{p}^{0}_{i})^{2}}\right)
\end{eqnarray}
we obtain that the net effect of the $z_N$-prescription is captured by the replacement
\begin{align}
	\label{eq:zN_to_Plus_prescr}
	\frac{1}{z_N}\to \left(\frac{1}{z}\right)_+-\frac{1}{2}\log\left( \frac{{\ktness}^2D^{2}}{4(\tilde{p}^{0}_{i})^{2}}\right)\delta(z)\,.
\end{align}
By summing over all possible splittings $a_i \to (*)$, we find that the differential jet functions in the $z_N$-prescription are
\begin{align}
&{\tilde \jmath}^{\,ss'}_{a_i}(\ktness/Q) = \sum_{(*)} {\tilde \jmath}^{\,ss'}_{a_i \to (*)}(\ktness/Q)
	= 2d_{a_i}^{ss'}  \,\frac{\left(\frac{\mu_R^2}{{\ktness}^2D^{2}}\right)^{\epsilon}}{\ktness/Q}\frac{e^{\epsilon\gamma_E}}{\Gamma(1-\epsilon)} \frac{\as}{\pi} \notag\\
	&\!\!\times \begin{cases}
		-\gamma_q(1-2\epsilon  \log 2) + C_F \left[-\frac{1}{2} \log \left( \frac{{\ktness}^2D^{2}}{4(\tilde{p}^{0}_{i})^{2}}\right)+\epsilon  \left(\frac{\pi ^2}{6}-\frac{7}{4}\right)\right] + O(\epsilon^2) &a_i=q \\
		-\gamma_g(1-2\epsilon  \log 2)+C_A\left[-\frac{1}{2} \log \left( \frac{{\ktness}^2D^{2}}{4(\tilde{p}^{0}_{i})^2}\right)+\epsilon \left(\frac{\pi^2}{6}-\frac{131}{72} \right) \right]
		+n_fT_R\epsilon\frac{17}{36}+ O(\epsilon^2)  &a_i=g
	\end{cases}
\end{align}
and the cumulant jet functions are
\begin{align}
	&\widetilde{\mathcal{J}}_{a_i}^{ss'}(\rcut) = d_{a_i}^{ss'}\, \frac{e^{\epsilon\gamma_E}}{\Gamma(1-\epsilon)}\left(\frac{\mu_R^2}{\rcut^2 D^2 Q^2}\right)^\epsilon \frac{\as}{\pi}  \notag \\
	&\times \begin{cases}
		\frac{C_{F}}{2\epsilon^2}+\frac{\gamma_q+C_{F}\log\left( \frac{\rcut D Q}{2\tilde{p}^{0}_{i}}\right)}{\epsilon}-2\gamma_q\log{2}+C_{F}\left(\frac{7}{4}-\frac{\pi^2}{6}\right)+O(\epsilon)&a_i=q \\
		\frac{C_{A}}{2\epsilon^2}+\frac{\gamma_g+C_{A}\log\left( \frac{\rcut D Q}{2\tilde{p}^{0}_{i}}\right)}{\epsilon}-2\gamma_g\log{2}+C_{A}\left(\frac{131}{72}-\frac{\pi^2}{6}\right)+n_{f}T_{R}\frac{17}{36}+O(\epsilon)&a_i=g\,\,\,.
	\end{cases}
\end{align}
By comparing the SCET-like jet functions regularised with the $z_{N}$-prescription with the result in Eq.~(\ref{eq:cumjetktness}), we find that, up to ${\cal O}(\epsilon^0)$, they differ only by the finite contribution
\begin{equation}
	\mathcal{J}_{a_i}^{ss'}(\rcut) - \mathcal{\widetilde J}_{a_i}^{ss'}(\rcut) = - d_{a_i}^{ss'}\, C_{a_i}\frac{\as}{\pi} \left(\frac{\pi^2}{12}+O(\epsilon) \right)\, .
\end{equation}

\bibliography{biblio}

\end{document}